\documentclass[fleqn,usenatbib]{mnras}

\usepackage{newtxtext,newtxmath}
\usepackage[T1]{fontenc}
\usepackage{ae,aecompl}
\usepackage{longtable}

\usepackage[english]{babel}
\usepackage{graphicx}
\usepackage{xcolor}

\newcommand{\degsq}{\mbox{$\rm deg^{2}$}}

\title[Orphan Afterglow AT\,2019pim]{The Luminous, Slow-Rising Orphan Afterglow AT2019pim as a Candidate Moderately Relativistic Outflow}

\newcommand{\ljmu}{1}
\newcommand{\cornell}{2}
\newcommand{\ttu}{3}
\newcommand{\caltech}{4}
\newcommand{\gsfc}{5}
\newcommand{\umd}{6}
\newcommand{\jssi}{7}
\newcommand{\ngf}{8}
\newcommand{\uw}{9}
\newcommand{\iit}{10}
\newcommand{\berkeley}{11}
\newcommand{\lsu}{12}
\newcommand{\ioffe}{13}
\newcommand{\usra}{14}
\newcommand{\intpt}{15}
\newcommand{\mitk}{16}
\newcommand{\radboud}{17}
\newcommand{\okc}{18}
\newcommand{\cfa}{19}
\newcommand{\ipac}{20}
\newcommand{\nrao}{21}
\newcommand{\mitpk}{22}
\newcommand{\coo}{23}
\newcommand{\miller}{24}

\author[Perley et al.]{Daniel A. Perley$^{\ljmu}$, %et al. 
Anna Y. Q. Ho$^{\cornell}$, 
Michael Fausnaugh$^{\ttu}$,
Gavin P. Lamb$^{\ljmu}$,
\newauthor
Mansi M. Kasliwal$^{\caltech}$,
Tomas Ahumada$^{\caltech}$,
Shreya Anand$^{\caltech}$,
Igor Andreoni$^{\gsfc,\umd,\jssi,\ngf}$,
\newauthor
Eric Bellm$^{\uw}$,
Varun Bhalerao$^{\iit}$,
Bryce Bolin$^{\gsfc}$,
Thomas G. Brink$^{\berkeley}$,
Eric Burns$^{\lsu}$,
\newauthor
S. Bradley Cenko$^{\gsfc,\jssi}$,
Alessandra Corsi$^{\ttu}$,
Alexei V. Filippenko$^{\berkeley}$,
Dmitry Frederiks$^{\ioffe}$,
\newauthor
Adam Goldstein$^{\usra}$,
Rachel Hamburg$^{\intpt}$, % 0000-0003-0761-6388
Rahul Jayaraman$^{\mitk}$,
Peter G. Jonker$^{\radboud}$,
\newauthor
Erik C. Kool$^{\okc}$,
Shrinivas R. Kulkarni$^{\caltech}$,
Harsh Kumar$^{\cfa,\iit}$,
Russ Laher$^{\ipac}$,
\newauthor
Andrew Levan$^{\radboud}$,
Alexandra Lysenko$^{\ioffe}$,
Richard A. Perley$^{\nrao}$,
George R. Ricker$^{\mitpk}$,
\newauthor
Reed Riddle$^{\caltech,\coo}$,
Anna Ridnaia$^{\ioffe}$,
Ben Rusholme$^{\ipac}$,
Roger Smith$^{\coo}$,
Dmitry Svinkin$^{\ioffe}$,     
\newauthor
Mikhail Ulanov$^{\ioffe}$,
Roland Vanderspek$^{\mitk}$,
Gaurav Waratkar$^{\iit}$,
Yuhan Yao$^{\berkeley,\miller}$
\\
\\
$^{\ljmu}$ Astrophysics Research Institute, Liverpool John Moores University, 146 Brownlow Hill, Liverpool L3 5RF, UK \\
$^{\cornell}$ Department of Astronomy, Cornell University, Ithaca, NY 14853, USA \\
$^{\ttu}$ Department of Physics and Astronomy, Texas Tech University, Box 1051, Lubbock, TX 79409-1051, USA \\
$^{\caltech}$ Cahill Center for Astrophysics, California Institute of Technology, MC 249-17, 1200 E. California Boulevard, Pasadena, CA 91125, USA \\
$^{\gsfc}$ Astrophysics Science Division, NASA Goddard Space Flight Center, Mail Code 661, Greenbelt, MD 20771, USA \\
$^{\umd}$ Department of Astronomy, University of Maryland, College Park, MD 20742, USA \\
$^{\jssi}$ Joint Space-Science Institute, University of Maryland, College Park, MD 20742, USA \\
$^{\ngf}$ Neil Gehrels Fellow \\
$^{\uw}$ DIRAC Institute, Department of Astronomy, University of Washington, 3910 15th Avenue NE, Seattle, WA 98195, USA \\
$^{\iit}$ {Physics Department, Indian Institute of Technology Bombay, Powai, 400 076, India} \\
$^{\berkeley}$ Department of Astronomy, University of California, Berkeley, CA 94720-3411, USA \\
$^{\lsu}$ Department of Physics \& Astronomy, Louisiana State University, Baton Rouge, LA 70803, USA \\
$^{\ioffe}$ Ioffe Institute, Polytekhnicheskaya, 26, St. Petersburg, 194021, Russia \\
$^{\usra}$ Science and Technology Institute, Universities Space Research Association, Huntsville, AL 35805, USA \\
$^{\intpt}$ Universit\'e Paris-Saclay, CNRS/IN2P3, IJCLab, 91405 Orsay, France \\
$^{\mitk}$  Kavli Institute for Astrophysics and Space Research, Massachusetts Institute of Technology, Cambridge, MA 02139, USA \\
$^{\radboud}$ Department of Astrophysics/IMAPP, Radboud University, P.O. Box 9010, 6500 GL Nijmegen, The Netherlands \\
$^{\okc}$ The Oskar Klein Centre, Department of Astronomy, Stockholm University, AlbaNova, SE-10691, Stockholm, Sweden \\
$^{\cfa}$ {Center for Astrophysics | Harvard \& Smithsonian, 60 Garden St. Cambridge MA, 02138, USA} \\
$^{\ipac}$ IPAC, California Institute of Technology, 1200 E. California Blvd., Pasadena, CA 91125, USA \\
$^{\nrao}$ National Radio Astronomy Observatory, PO Box 0, Socorro, NM 87801, USA \\
$^{\mitpk}$ Department of Physics, and Kavli Institute for Astrophysics and Space Research, Massachusetts Institute of Technology, Cambridge, MA 02139, USA\\
$^{\coo}$ Caltech Optical Observatories, California Institute of Technology, Pasadena, CA  91125 \\
$^{\miller}$ Miller Institute for Basic Research in Science, 468 Donner Lab, University of California, Berkeley, CA 94720, USA \\
}

\date{Accepted 08 Jan 2025. Received YYY; in original form ZZZ}

\pubyear{2025}

\begin{document}
\label{firstpage}

\maketitle

\clearpage

\begin{abstract}
Classical gamma-ray bursts (GRBs) have two distinct emission episodes: prompt emission from ultrarelativistic ejecta and afterglow from shocked circumstellar material.  While both components are extremely luminous in known GRBs, a variety of scenarios predict the existence of luminous afterglow emission with little or no associated high-energy prompt emission.  We present AT\,2019pim, the first spectroscopically confirmed afterglow with no observed high-energy emission to be identified.  Serendipitously discovered during follow-up observations of a gravitational-wave trigger and located in a contemporaneous TESS sector, it is hallmarked by a fast-rising ($t \approx 2$\,hr), luminous ($M_{\rm UV,peak} \approx -24.4$\,mag) optical transient with accompanying luminous X-ray and radio emission.  No gamma-ray emission consistent with the time and location of the transient was detected by {\it Fermi}-GBM or by {\it Konus}, placing constraining limits on an accompanying GRB.  We investigate several independent observational aspects of the afterglow in the context of constraints on relativistic motion and find all of them are consistent with an initial Lorentz factor of $\Gamma_0 \approx$\ 10--30 for the on-axis material, significantly lower than in any well-observed GRB and consistent with the theoretically predicted ``dirty fireball'' scenario in which the high-energy prompt emission is stifled by pair production.  However, we cannot rule out a structured jet model in which only the line-of-sight material was ejected at low-$\Gamma$, off-axis from a classical high-$\Gamma$ jet core, and an on-axis GRB with below-average gamma-ray efficiency also remains a possibility.
This event represents a milestone in orphan afterglow searches, demonstrating that luminous optical afterglows lacking detected GRB counterparts can be identified and spectroscopically confirmed in real time.
\end{abstract}


\section{Introduction}

Long-duration gamma-ray bursts (GRBs) originate from the collapse of a rapidly rotating, stripped-envelope massive star.  During the collapse, both a highly collimated relativistic jet and a largely isotropic supernova (SN) explosion are produced; the collision of the jet with the surrounding medium also produces a multiwavelength afterglow (for reviews see, e.g., \citealt{vanParadijs2000,Piran2004,Woosley+2006,Hjorth+2012,Gehrels+2012}).   

The properties of the SN show little variation from event to event \citep{Cano+2014,Cano+2017,Melandri+2014}.  All known GRB-associated supernovae (SNe) are of spectral type Ic-BL; the SN peak luminosity varies by only about a factor of 2--3 and the rise time varies even less, suggesting a common progenitor with relatively little intrinsic diversity in (for example) structure or composition.\footnote{Known exceptions are plausibly associated with other classes of events: a few GRBs with $t_{90}>2$\,s but with strong upper limits on an accompanying classical SN (GRBs 060605, 060614, 211211A, and 230307A; \citealt{Fynbo+2006,Gehrels+2006,DellaValle+2006,Rastinejad+2022,Troja+2022,Levan+2023}) may be related to short-duration GRBs or perhaps another class of event entirely \citep{GalYam+2006,Ofek+2007,Zhang+2007,Jin+2015,Yang+2022,Yang+2024}, while GRB 111209A and its unusually luminous SN is a member of the ultra-long class of GRBs \citep{Greiner+2015,Levan+2014,Gendre+2013}.}

The nature of the jet, however, is vastly more diverse.  Inferred GRB isotropic-equivalent gamma-ray energies ($E_{\gamma, {\rm iso}}$) vary from $10^{46}$ to almost $10^{55}$\,erg, while the duration, spectral hardness, and temporal structure of the GRB light curve also vary greatly \citep{Kouveliotou+1993,Paciesas+1999,Amati2006}.  Some of this variation may originate from simple differences in orientation angle (a ``structured jet''; e.g., \citealt{Meszaros+1998,Dai+2001,Lipunov+2001,Rossi+2002,Zhang+2002,Granot+2003}), although to what extent intrinsic versus viewing angle effects govern the observed diversity remains a subject of debate (e.g., \citealt{Kulkarni+1998,Soderberg+2004,Lamb+2005,Amati+2007,Cenko+2010,Cenko+2011,Pescalli+2015,Salafia+2015,Salafia+2020,Beniamini+2020a,Salafia+2022,OConnor+2023}).

GRBs are by definition selected at high photon energies ($>$10~keV) via the detection of prompt emission by an orbiting wide-field-of-view satellite, 
 which is then followed by a narrow-field search for the associated afterglow and/or SN.  However, there is no strong reason to expect that all energetic jet outflows must produce luminous gamma-ray emission of this nature.   The outflow might, for example, be insufficiently variable to generate the luminous internal shocks that are generally presumed to produce GRB prompt emission \citep{Rees+1994}.   Alternatively, the velocity of the ejecta may be sufficiently low that pair production suppresses the production of the highest-energy photons (a ``dirty'' fireball; \citealt{Dermer+2000,Huang+2002,Rhoads2003}).  Geometrical reasons may also be important: the GRB ejecta that produce the prompt emission travel much faster (and beam radiation into a narrower opening angle) than the afterglow, which by definition is only set up once the outflow has decelerated somewhat \citep{Rhoads1997,Perna+1998,Nakar+2002,Granot+2002,Rhoads2003}.  The rate of these various types of gamma-ray-``dark'' explosions may greatly exceed that of classical long-duration GRBs.

Finding examples has, however, proven quite challenging.  The optical, X-ray, and radio sky are all much more crowded than the gamma-ray sky, requiring the advent of both wide-field telescopes and sophisticated machine-learning techniques to distinguish genuine transients.   There are also many false positives with similar ``fast-rise, slow-decay'' features.   In the optical band, flares from M-dwarfs and cataclysmic variables (dwarf novae) are particularly problematic \citep{Kulkarni2006,Rau2008,Berger2013,Ho2018,Andreoni+2020}: at typical operational flux limits, the rates of these events exceed the expected rate of afterglows by orders of magnitude.

However, the past ten years have seen steady progress.  The first optical orphan\footnote{We use the term ``orphan'' to describe any afterglow without an observationally associated GRB.} afterglow candidate was PTF11agg \citep{Cenko+2013}, found by the Palomar Transient Factory during a dedicated, high-cadence, narrow-field experiment.   PTF11agg was detected as a new, bright ($r \lesssim 18.25$\,mag) transient in the first exposure of the field taken that night, and faded rapidly in subsequent exposures over the next few hours.   Follow-up observations with the Karl G. Jansky Very Large Array (VLA; \citealt{PerleyR+2011}) revealed a long-lived scintillating radio counterpart; deep late-time optical imaging after the optical transient faded unveiled a faint, blue extended object at the location --- most likely a high-redshift ($z\gtrsim0.5$) host galaxy, although its actual redshift remains unknown and its cosmological nature unconfirmed.
Unfortunately, because of poor constraints on the true explosion time (a window of 20\,hr between the most recent limit and first detection), it was not possible to rule out that the location was in a ``blind spot'' to {\it Fermi} and/or other satellites at the time of explosion.

Two other optical afterglows were subsequently discovered by wide-area sky surveys in a similar manner: iPTF14yb \citep{Cenko+2015} and ATLAS17aeu \citep{Stalder+2017,Bhalerao+2017,Melandri2019}.  iPTF14yb was spectroscopically confirmed to originate at a cosmological distance (redshift $z=1.9733$).  \mbox{ATLAS17aeu}, discovered serendipitously in follow-up observations of a gravitational-wave trigger, also likely originated at $z>1$ given the photometric properties of its presumptive host galaxy, although (as with PTF11agg) it has not been possible to confirm this spectroscopically.  However, both events were later found to have associated GRBs detected by {\it Fermi} or other satellites whose times and sky locations were consistent with the optically discovered afterglows.  

Dedicated afterglow searches with the Zwicky Transient Facility (ZTF) have yielded nine published\footnote{This total does not include several other events distributed via GCN Circulars but not yet published, including the notable events AT\,2023lcr \citep{AN2023.178} and AT\,2023sva \citep{AN2023.251}.} afterglow candidates to date \citep{Ho2020,Ho2022,Andreoni+2021,Andreoni+2022}, of which seven have redshift measurements from optical spectroscopy. Redshifts range from $z=0.876$ (AT\,2021buv; \citealt{Ho2022}) to $z=2.9$ (AT\,2020blt; \citealt{Ho2020}).
Of the nine events, three had no associated detected GRB (AT\,2020blt, AT\,2021any, and AT\,2021lfa).  However, at the redshifts of these three events an accompanying typical GRB cannot be ruled out based on the sensitivity and coverage of GRB satellites (\citealt{Ho2020,Ho2022}; but see  \citealt{Lipunov2022}, who refine the explosion time of AT\,2021lfa and present deeper limits on gamma-ray emission that are more constraining).
As a result, it is unclear from the ZTF observations alone if these objects represent normal GRBs whose prompt high-energy emission was simply missed.  Modeling the X-ray through radio emission, and the detection of a ``rise phase'' using the MASTER telescope network, has led to suggestions that at least some of these events had a truly low Lorentz factor \citep{Lipunov2022,Xu2023}, and that another may represent a GRB with a low gamma-ray efficiency \citep{Sarin2022}. 

Advancements have also been made outside the optical domain.  An X-ray transient with GRB-like properties and no known GRB counterpart was reported by \cite{Bauer+2017}, although it is much lower in luminosity than classical GRB afterglows and is also spectroscopically unconfirmed 
(but is convincingly associated with a high-redshift galaxy with photo-$z$ of $2.23^{+0.98}_{-1.84}$).
Separately, searches for orphan afterglows using radio-survey data have identified a compelling candidate radio afterglow, plausibly from a highly off-axis GRB \citep{Law+2018}, but the explosion time window is years long and it is not possible rule out a classical GRB origin.  Additionally, radio follow-up observations of optically discovered SNe~Ic-BL have identified a few with moderately luminous radio emission indicative of a very energetic high-velocity shock, although no clear evidence of a jetted, relativistic outflow has yet emerged \citep{Soderberg+2010,Margutti+2014,Corsi+2017,Marongiu+2019}.

In this paper we describe the discovery of AT\,2019pim (ZTF19abvizsw), the first unambiguous optical afterglow of a relativistic explosion with secure limits on accompanying GRB-like high-energy emission.
In \S 2 we briefly outline the ZTF afterglow search program and the partially serendipitous discovery of AT\,2019pim during a gravitational-wave counterpart search, and describe our observational follow-up activities that confirmed this source as an afterglow.  We model the observational properties in \S 3, including the explosion time and peak time using a combination of our ZTF discovery observations, follow-up observations, and TESS data, and we place upper limits on associated gamma-ray emission from {\it Konus} and {\it Fermi}. \S 4 establishes physical constraints on the nature of the outflow using the combined optical and radio data set, and we summarise our conclusions in \S 5.

\section{Observations}

\subsection{P48 Discovery}

The Zwicky Transient Facility (ZTF; \citealt{Bellm+2019a,Graham+2019}) is a refurbishment of the Palomar 48-inch Oschin Schmidt telescope (P48), most recently in use as part of the Palomar Transient Factory (PTF).  The ZTF camera has a 47 square degree operational field of view, fast readout, and near-real-time data processing \citep{Masci+2019,Dekany+2020}.

A major science driver of ZTF has been the search for luminous, fast, and/or young transients (characteristic timescales $<1$\,day).  While such transients can be detected in the standard 2--3\,d cadence public survey \citep{Bellm+2019b}, higher-cadence observations (nightly or faster cadence) are better suited for rapid and accurate identification of these objects.  Several ZTF programmes operate at higher cadence, including a 10,000 square degree ``partnership'' survey which acquires 4 observations of each field per night and a 2000--3000\,\degsq 1-night cadence survey.  During 2019 and 2020, ZTF also conducted a public 1-night cadence survey shadowing the Transiting Exoplanets Survey Satellite (TESS; \citealt{Ricker+2015}) footprint \citep{vanRoestel+2019}.  Custom software filters scan all of these streams to search for bright new transients not coincident with known point sources.

%2,458,728.47988 JD
On 2019 September 1 (23:31:01.838 UTC\footnote{UT dates are used throughout this paper.}; equivalent to 58727.97988 MJD) the Laser Interferometer Gravitational-wave Observatory (LIGO) - Virgo Gravitational Wave Interferometer (Virgo) network \citep{LIGO,Acernese+2015} registered a candidate gravitational-wave signal, initially designated S190901ap \citep{GCN25606}, consistent with a neutron-star--neutron-star merger waveform\footnote{The astrophysical nature of this event has not been confirmed by further analysis \citep{Abbott+2021}.}.  Only LIGO L1/Livingston detected the event (H1/Hanford was offline) and thus the localisation was exceptionally poor, covering over 14,000\,deg$^2$; the distance constraint is 241 $\pm$ 79\,Mpc ($z=0.054\pm0.017$)\footnote{We assume $\Omega_M=0.3$, $\Omega_\Lambda=0.7$, $h=0.7$ throughout this work.}.  Nevertheless, ZTF was triggered in target-of-opportunity mode for the following night to tile as much of the observable error region as possible and all candidates detected during the night with no previous history were scanned by eye using tools available via the GROWTH Marshal \citep{Kasliwal+2019}.

\begin{figure*}
\includegraphics[width=\textwidth]{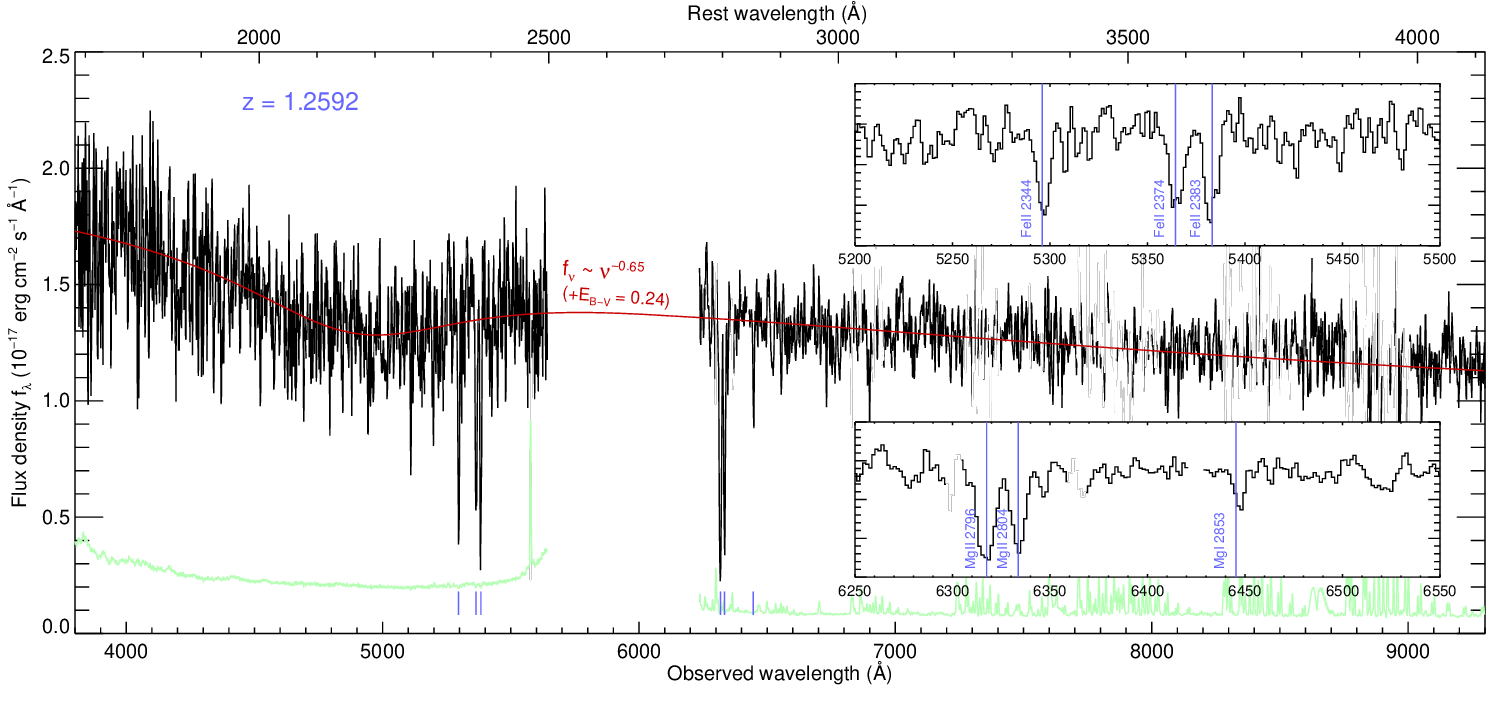}
\caption{LRIS spectrum of AT\,2019pim.  The spectrum has been lightly smoothed by convolution, and regions affected by strong night-sky lines are plotted in grey.   A continuum model is overplotted in red: this is a power-law ($F_\nu \propto \nu^{-0.65}$) extinguished by host-galaxy dust ($E_{B-V}=0.24$\,mag, using a \citealt{Fitzpatrick1999} dust model with $R_V = 3.1$ and $c_3=1.0$) and Galactic dust ($E_{B-V}=0.038$\,mag).   Inset panels show zoom-ins on two strong-line regions: the Fe~II series (upper panel) and the Mg~II\,/\,Mg~I series (lower panel).  The error spectrum (after convolution) is shown in light green.}
\label{fig:spectrum}
\end{figure*}

AT\,2019pim was first detected in ZTF data at MJD = 58728.1798 %JD=2458728.6798 
with a magnitude\footnote{Magnitudes are in the AB system \citep{Oke+1983} and uncertainties are 1$\sigma$ throughout, unless otherwise specified.} of $g=20.04 \pm 0.16$. 
Following a filter change, the source was detected again $\sim 1$\,hr later (MJD = 58728.2300) with $r=19.45\pm0.11$\,mag at a consistent location ($\alpha = 18^{\rm h}37^{\rm m}53.48^{\rm s}$, $\delta = +61^\circ 29' 52.74''$; J2000). 
There is no counterpart in prior ZTF reference imaging nor any previous detections of variability at the same location.  The most recent nondetection originates from the preceding night (5$\sigma$ limit of $g>20.60$\,mag at MJD = 58727.3161).  The source was within the TESS footprint and the associated footprint of the public ZTF 1-night TESS shadowing survey, but the alerts it generated did not enter the public stream because the gravitational wave target-of-opportunity search programme displaced normal public observations that night.  It 
passed an automated software filter designed to find young SNe and was ``saved'' (i.e., flagged as a transient of interest) after scanning the output of this and other filters for candidate counterparts of the GW event.   The transient was reported to the GCN Circulars (as ZTF19abvizsw, its internal ZTF survey name), along with the three other transients detected that night consistent within the error region with no prior history \citep{GCN25616}.  All four candidates were reported to the Transient Name Server the next day \citep{TNSTR1722}.

\subsection{Spectroscopy}

On the night following the discovery of the transient \mbox{(2019-09-03)} we obtained a spectrum using the Low Resolution Imaging Spectrometer (LRIS; \citealt{Oke+1995}) on the Keck I 10\,m telescope.  The 600/4000 grism was used on the blue side and the 600/7500 grating was used on the red side, providing wavelength coverage of 3139--5642\,\AA\ (blue) and 6236--9516\,\AA\ (red).  The $1''$  slit was used, positioned at the parallactic angle (134\,$\deg$ at the time of observation).  The exposure time was 600\,s on both sides.  The spectrum was reduced using LPipe \citep{Perley2019} with BD+284211 as a flux calibrator.  The red and blue relative flux scales are scaled by matching synthetic photometry to colours inferred from photometry of the transient.

The reduced spectrum, shown in Figure \ref{fig:spectrum}, is largely featureless and fairly red.  Deep, narrow absorption lines are evident in the middle region of the spectrum; these are matched by Fe~II, Mg~II, and Mg~I at a common redshift of $z = 1.2592$ $\pm$ 0.0004.
Because the signal-to-noise ratio (S/N) of the spectrum does not permit the detection of fine-structure lines, this is technically only a lower limit on the true redshift.  A firm upper limit of $z<2.2$ can be placed by the absence of Lyman $\alpha$ at $\lambda>3900$\,\AA, where the S/N of the spectrum is relatively high.   In spite of this, we can be reasonably confident that the absorption redshift is indeed that of AT2019pim: the strength of the absorption lines (in particular of the MgII~2796 line, for which we measure a rest-frame equivalent width of $W_{r}$ = 4.0~$\pm$~0.3 \AA) is much higher than in typical line-of-sight absorbers \citep{Christensen+2017,Churchill+2020}, and our spectrum rules out any strong ($W_{r} \gtrsim 1$\,\AA) higher-redshift Mg\,II absorption system between $1.26<z<2.2$.  We will assume $z=1.2592$ throughout this work.

The implied rest-frame UV magnitude (AB) at the time of the $g$-band discovery is $M_{2170\rm\,\mathring{A}}=-24.4$ (for $z=1.2596$, as will be assumed throughout the remainder of this paper).  This unambiguously identifies the event as an extremely luminous cosmological explosion and (given the inconsistent distances) firmly rules out any association with the gravitational-wave trigger.  

\subsection{Follow-up Photometry}

We used several different telescopes at locations around the globe to obtain additional photometric observations of AT\,2019pim over the first few nights following its discovery.  These include the GROWTH India Telescope (GIT; \citealt{Kumar+2022}), the Liverpool Telescope (LT), and the Apache Point Observatory 3\,m telescope (APO).  We additionally acquired later imaging observations of the transient with ACAM on the William Herschel Telescope (2019-09-11/12), with LRIS on the Keck I 10\,m telescope (2019-09-24 and 2019-10-27), and with OSIRIS on the GTC (2019-11-23).  Late-time reference imaging of the host galaxy was taken with LRIS in April 2022 using the LRIS $U$, $G$, $R$, and $RG850$ filters.

\begin{table}
	\centering
	\caption{Host photometry}
	\label{tab:hostphot}
	\begin{tabular}{ll}
		\hline
		filter & AB magnitude\\
		\hline
         $u$ &  24.63 $\pm$ 0.09 \\
         $g$ &  24.29 $\pm$ 0.07 \\
         $r$ &  24.50 $\pm$ 0.15 \\
         $R$ &  24.20 $\pm$ 0.06 \\
         $i$ & (24.08)$^{\dagger}$ \\
         $z$ &  23.38 $\pm$ 0.09 \\
		\hline
	\end{tabular}
\newline $^{\dagger}$Estimated using SED fitting (\S \ref{sec:hostsedfit})
\end{table}

Photometry for most follow-up observations was performed using a custom aperture photometry routine in IDL, with calibration performed relative to Sloan Digital Sky Survey (SDSS) secondary standard stars in the field.  

The host galaxy of this source is relatively bright and contributes non-negligibly to the flux at all epochs.  For most of our measurements, we correct for the host contribution in flux space by measuring the host flux in the late-time LRIS imaging ($U$, $G$, $R$, and $RG850$ were treated as $u_{\rm SDSS}$, $g_{\rm SDSS}$, $R_C$, and $z_{\rm SDSS}$, respectively) and GTC imaging\footnote{The GTC observations were taken 83\,days post-explosion, when afterglow contribution may still have been present.  Our empirical model (\S\ref{sec:lateoptical}) suggests that the afterglow had $r \approx 27.5$\,mag at this time, which would represent about  0.06\,mag contribution to the host measurement.  This is less than the 1$\sigma$ statistical uncertainty in the photometry, and we did not correct for this in our estimate in Table \ref{tab:hostphot}.} ($r_{\rm SDSS}$) and subtracting the fluxes from the direct aperture photometry measurements.  
No reference imaging was acquired in the $i$ band, so the host flux at this band was inferred indirectly via synthetic photometry of our fit to the host spectral energy distribution (SED).  The host-galaxy magnitudes are given in Table \ref{tab:hostphot}.

While the host galaxy is compact and direct flux subtraction should generally be adequate, in the case of the LRIS measurements in September and October we employ image subtraction to obtain the flux of the afterglow above the level of the host galaxy.  This was not possible for the simultaneous LRIS $i$-band observations, since no late-time reference image was obtained in this band.   The last epoch resulted in nondetections in both bands; upper limits are for an aperture fixed at the afterglow location and given as 2.5$\sigma$.
Photometry is presented in Table \ref{tab:phot}.

\begin{table}
	\centering
	\caption{Ground-based Photometry of AT\,2019pim}
	\label{tab:phot}
	\begin{tabular}{llcrl}
		\hline
facility & MJD &  filter & AB mag. & unc. \\
		\hline
P48+ZTF    & 58727.1641 & g &$>$20.74 & $^{a}$   \\
P48+ZTF    & 58727.1790 & g &$>$20.77 & $^{a}$   \\
P48+ZTF    & 58727.2708 & g &$>$20.64 & $^{a}$   \\
P48+ZTF    & 58727.2933 & r &$>$20.49 & $^{a}$   \\
P48+ZTF    & 58727.3161 & g &$>$20.60 & $^{a}$   \\
P48+ZTF    & 58728.1798 & g & 20.04 & 0.16 \\
P48+ZTF    & 58728.2297 & r & 19.45 & 0.11 \\
GIT        & 58728.6034 & r & 20.34 & 0.09 \\
GIT        & 58728.6100 & r & 20.35 & 0.06 \\
GIT        & 58728.6189 & i & 20.12 & 0.09 \\
GIT        & 58728.6259 & i & 19.97 & 0.08 \\
GIT        & 58728.6353 & r & 20.40 & 0.08 \\
GIT        & 58728.7223 & i & 19.96 & 0.08 \\
GIT        & 58728.7287 & i & 20.09 & 0.09 \\
GIT        & 58728.8026 & g & 21.23 & 0.11 \\
GIT        & 58728.8093 & g & 21.07 & 0.10 \\
LT+IOO     & 58729.8552 & r & 21.54 & 0.18 \\
LT+IOO     & 58729.9420 & r & 21.70 & 0.09 \\
LT+IOO     & 58729.9481 & g & 22.16 & 0.10 \\
LT+IOO     & 58729.9541 & i & 21.24 & 0.07 \\
LT+IOO     & 58729.9621 & z & 20.96 & 0.12 \\
LT+IOO     & 58730.8802 & r & 22.17 & 0.10 \\
LT+IOO     & 58730.8862 & g & 22.73 & 0.21 \\
LT+IOO     & 58730.8923 & i & 21.72 & 0.10 \\
LT+IOO     & 58730.8983 & z & 21.33 & 0.19 \\
LT+IOO     & 58731.0112 & r & 22.23 & 0.12 \\
LT+IOO     & 58731.0172 & g & 22.78 & 0.17 \\
LT+IOO     & 58731.0232 & i & 21.76 & 0.11 \\
LT+IOO     & 58731.0293 & z & 21.75 & 0.22 \\
APO        & 58733.2200 & r & 22.36 & 0.03 \\
APO        & 58733.2530 & i & 21.94 & 0.05 \\
APO        & 58733.2650 & g & 22.78 & 0.11 \\
WHT+ACAM   & 58737.8819 & r & 22.77 & 0.10 \\
WHT+ACAM   & 58738.8839 & i & 22.55 & 0.09 \\
WHT+ACAM   & 58739.9939 & i & 22.67 & 0.13 \\
WHT+ACAM   & 58740.0219 & r & 23.48 & 0.27 \\
Keck1+LRIS & 58750.2351 & g & 25.44 & 0.36 \\
Keck1+LRIS & 58750.2354 & i & 25.50 & 0.63 \\
Keck1+LRIS & 58783.2306 & i &$>$24.66 & $^{b}$   \\
Keck1+LRIS & 58783.2306 & g &$>$26.22 & $^{b}$   \\
		\hline
	\end{tabular}
{\par \begin{flushleft}
$^{a}$ ZTF limits are 5$\sigma$ alert-photometry limits for the associated image. \\
$^{b}$ LRIS limits are 2.5$\sigma$ forced photomerty limits at the source location.
\end{flushleft}}
\end{table}

\begin{figure}
\includegraphics[width=0.47\textwidth]{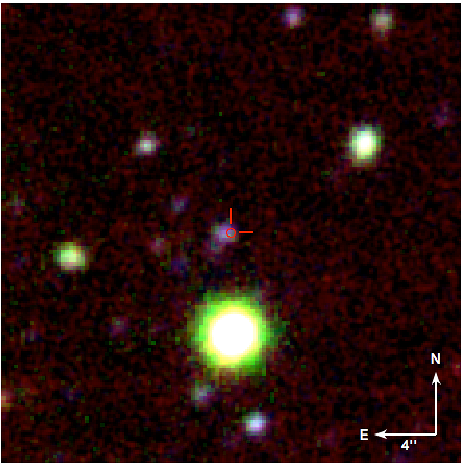}
\caption{Late-time imaging from LRIS ($g$, $i$) and GTC ($r$), combined into a false-colour image.  The image is 30$^{\prime \prime}$ across.  The afterglow location, shown at centre in red, is coincident with a blue, extended source, also seen in (shallower) Legacy Survey imaging of the field.
}
\label{fig:host}
\end{figure}

\subsection{TESS observations}

As previously noted, AT\,2019pim was detected in a high-cadence ZTF field associated with an active TESS sector.  TESS observed the field nearly continuously during Sector 15 from 2019-08-15 to 2019-09-10 in Camera 2, CCD 2.  The location was imaged in nearly 2400 30\,min full-frame images (FFIs) over that period.

\begin{figure*}
\includegraphics[width=0.97\textwidth]{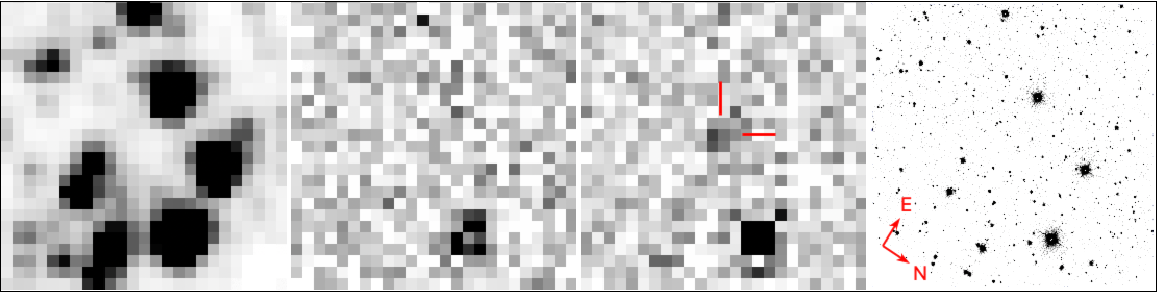}
\caption{TESS imaging of AT\,2019pim.  The left panel shows a stacked pre-explosion image of the field.
The two middle panels display individual FFIs after subtraction of this reference.  The middle-left panel is from an FFI taken at TJD 1728.5004505, $\sim 0.1$\,d before the inferred onset time; the detected object near the bottom is a bright variable star in the field.  The middle-right panel is from an FFI taken at peak (TJD 1728.667117), showing the detection of the afterglow (marked).   A Liverpool Telescope image is shown at right for reference.  Images are $8.3'$ on each side.}
\label{fig:tessim}
\end{figure*}
% invert X and rotate 125 degrees to match TESS to cardinal
% 21 arcsec/pix 

A light curve for AT\,2019pim was constructed from the FFIs using difference imaging.  First, we constructed a reference image by median stacking 20 FFIs with low background levels (Figure \ref{fig:tessim}).  We then subtracted the reference image from each epoch using the ISIS software \citep{Alard+1998,Alard2000}, which solves for a spatially variable kernel that matches the point-spread function (PSF) of the reference image to individual FFIs.  This procedure removes systematic errors due to pointing shifts/jitter and thermal variations, and is able to recover clear but weak detections of the transient in individual images.  We extracted a light curve by fitting a model of the PSF to the difference images at the predicted location of the transient in the FFIs based on the coordinates of AT\,2019pim, and subtracted a local background based on the median of pixel values in an annulus of inner/outer radius 8/12 pixels, following similar procedures as by \cite{Fausnaugh+2021,Fausnaugh+2023}. 

Despite background subtraction and PSF-fitting, the long-term light curve shows slow ($\sim 1$\,d), low-level ($\sim 10$\,$\mu$Jy) variations in the baseline flux.   The origin of this is not completely certain, but is likely due to a combination of real variation in nearby bright stars that are blended with the transient and its background annulus, and (particularly in the days after the afterglow onset) variations in the background as the Earth limb becomes visible to the spacecraft. 
To model these background estimations we first subtracted a model of the late-time afterglow flux based on ground-based data (\S \ref{sec:lateoptical}), then measured the remaining background flux using a series of median windows with a duration of 0.5\,d each spanning from 3\,d prior to the likely explosion time to 3\,d after, excluding a 1\,d region around the afterglow onset.  A fourth-order polynomial was then fit to the median-averaged data, and the resulting background model was subtracted from the raw count values to estimate the afterglow count rate.  

The photon-counting uncertainties in the count measurements substantially underestimate the actual variation from exposure to exposure, even on short timescales when no background or afterglow variation is expected.   We calculated corrected errors by taking the standard deviation of the afterglow- and background-subtracted flux over each of the median windows described above and fit this with a second-order polynomial to model the time dependence of the noise.

TESS count values are converted to flux-density values (at the TESS central wavelength of $\lambda$ = 7865\,\AA) using a conversion factor of 0.01208 $\mu$Jy/count, calculated assuming a standard (Vega-like) spectrum and an on-source integration time of (1800\,s) $\times$ (0.8) $\times$ (0.99) = 1425.6\,s per FFI exposure.\footnote{The $\sim$20\% reduction in effective integration time is a consequence of the on-board cosmic ray excision procedure \citep{TESShandbook}.}

The TESS light curve is given in Table \ref{tab:tessphot}.  The counts column provides values prior to any background subtraction; the flux column lists values after background subtraction.  Observations taken more than 0.2\,d before or after the probable explosion time were binned together in proportion to the time before or after explosion.

\begin{table}
	\centering
	\caption{TESS photometry of AT\,2019pim near the time of outburst}
	\label{tab:tessphot}
	\begin{tabular}{lrrrr}
		\hline
MJD$^{a}$ & Counts$^{b}$ & $F_\nu^{c}$   & $\sigma^{d}$ & $n_{\rm FFI}^{e}$\\
    &        & ($\mu$Jy) & ($\mu$Jy) & \\
		\hline
58727.74062 &  -1215 &  -6.58 &  12.58 & 3 \\
58727.80312 &    207 &  11.25 &  12.63 & 3 \\
58727.85520 &  -1873 & -13.35 &  15.52 & 2 \\
58727.89688 &  -1425 &  -7.51 &  15.56 & 2 \\
58727.93854 &   -411 &   5.17 &  15.59 & 2 \\
58727.98020 &   -555 &   3.86 &  15.63 & 2 \\
58728.02188 &    260 &  14.12 &  15.67 & 2 \\
58728.05312 &  -1055 &  -1.45 &  22.20 & 1 \\
58728.07396 &    387 &  16.19 &  22.22 & 1 \\
58728.09479 &  -2752 & -21.54 &  22.25 & 1 \\
58728.11562 &   -524 &   5.59 &  22.28 & 1 \\
58728.13646 &    448 &  17.54 &  22.30 & 1 \\
58728.15729 &   3774 &  57.93 &  22.33 & 1 \\
58728.17812 &   9708 & 129.81 &  22.35 & 1 \\
58728.19895 &   8994 & 121.39 &  22.38 & 1 \\
58728.21979 &   4562 &  68.06 &  22.40 & 1 \\
58728.24062 &   5409 &  78.49 &  22.43 & 1 \\
58728.26145 &   6813 &  95.65 &  22.45 & 1 \\
58728.28229 &   6737 &  94.93 &  22.47 & 1 \\
58728.30312 &   7242 & 101.22 &  22.50 & 1 \\
58728.32395 &   6121 &  87.88 &  22.52 & 1 \\
58728.34479 &   5487 &  80.41 &  22.55 & 1 \\
58728.37603 &   3780 &  60.08 &  15.97 & 2 \\
58728.41770 &   4082 &  64.11 &  16.00 & 2 \\
58728.45937 &   2199 &  41.74 &  16.03 & 2 \\
58728.50103 &   4200 &  66.26 &  16.07 & 2 \\
58728.55312 &   4604 &  71.59 &  13.15 & 3 \\
58728.61562 &   3785 &  62.21 &  13.19 & 3 \\
58728.67812 &   1384 &  33.70 &  13.23 & 3 \\
58728.75451 &   1587 &  36.73 &  13.27 & 3 \\
58728.83437 &    305 &  21.79 &  11.54 & 4 \\
58728.92812 &    968 &  30.40 &  10.36 & 5 \\
58729.03228 &   1402 &  36.23 &  10.40 & 5 \\
58729.14687 &   -404 &  14.95 &   9.54 & 6 \\
		\hline
	\end{tabular}
{\par \begin{flushleft}
$^{a}$\,Midpoint of observation. \\
$^{b}$\,TESS counts, prior to subtraction of the time-dependent background model. (For binned rows, this is the mean counts per exposure.) \\
$^{c}$\,TESS flux density, after subtraction of the time-dependent background model.  Not corrected for Galactic or host extinction. \\
$^{d}$\,TESS flux-density uncertainty, based on the noise model. \\
$^{e}$\,Number of exposures (FFIs) binned together. \\
{\emph Note}: This table includes only measurements close to the inferred onset time of the afterglow.  A complete table of all TESS measurements with no binning applied is provided in the online supplementary material.
\end{flushleft}}
\end{table}

\subsection{Limits on a GRB Counterpart}
\label{sec:grblimits}

We searched the {\it Fermi}\footnote{\url{https://heasarc.gsfc.nasa.gov/W3Browse/fermi/fermigbrst.html}} \citep{Gruber2014,vonKienlin2014,Bhat2016},
{\it Fermi} subthreshold\footnote{\url{https://gcn.gsfc.nasa.gov/fermi\_gbm\_subthresh\_archive.html}} (with reliability flag \texttt{!=2}),
Swift\footnote{\url{https://swift.gsfc.nasa.gov/archive/grb\_table/}}, and General Coordinates Network\footnote{\url{https://gcn.gsfc.nasa.gov/gcn3\_archive.html}} archives for a GRB between the last ZTF nondetection and the first ZTF detection.
The only event that occurred during this period was the known GRB\,190901A,
at MJD 58727.89015.  The position of this GRB is inconsistent with that of AT\,2019pim and its time of occurrence was several hours before the optical explosion-time window (\S \ref{sec:explosiontime}), so an association can be firmly ruled out.

The position of AT\,2019pim was in the field of view of the {\it Fermi} Gamma-Ray Burst Monitor (GBM; \citealt{Meegan2009})
throughout the period between the most recent ZTF upper limit and ZTF discovery except for brief Earth occultations and South Atlantic Anomaly (SAA) passages.  We ran the GBM targeted search in the 10--1000\,keV energy band during this period.  The detector count data was separated into 1\,min blocks, each of which was analysed on 1\,s and 8\,s sliding time windows and, assuming a spectral model (described below), checked for detector-coherent flux above the background level.

\begin{figure}
\includegraphics[width=0.47\textwidth]{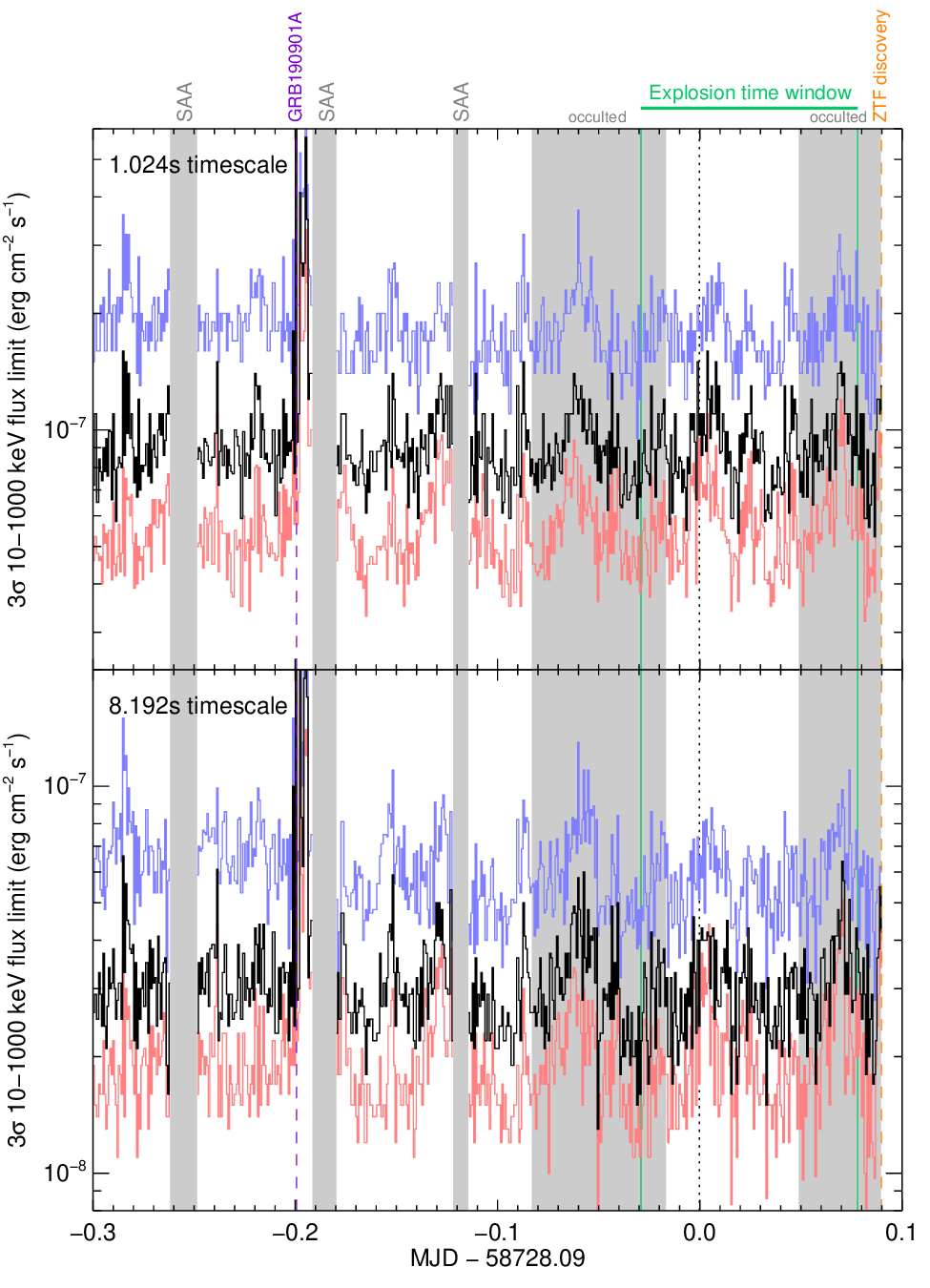}
\caption{Fermi GBM limits on gamma-ray emission (10--1000\,keV) near the time of onset of AT\,2019pim.   The black curve shows limits assuming a standard GRB spectrum of $E_{\rm peak} = 230$\,keV; the red curves assumes a soft spectrum ($E_{\rm peak} = 70$\,keV), and the blue curve assumes a hard spectrum ($E_{\rm peak} = 1500$\,keV).  The upper plot shows limits on the flux averaged over 1\,s intervals and the bottom plot over 8\,s intervals.  Limits are calculated in 1\,min windows and are 3$\sigma$.  SAA passages and occultations are indicated as shaded regions.  The explosion-time window as inferred from modeling of the optical rise is indicated (\S \ref{sec:explosiontime}).  
}
\label{fig:gbm-search}
\end{figure}

Limits were calculated for two different search timescales (1.024\,s and 8.192\,s) and three different spectral models, shown in Figure \ref{fig:gbm-search}.  Our preferred spectral model is parameterised using a \cite{Band2003} function with $E_{\rm peak} = 230$\,keV, $\alpha=-1.0$, $\beta=-2.3$, and is shown as a black curve, although for comparison we also provide limits assuming two other models: a ``soft'' model assuming a Band spectrum and $E_{\rm peak}=70$\,keV, $\alpha=-1.9$, $\beta=-3.7$, and a ``hard'' model with a cutoff power law \citep{Goldstein+2016} and $E_{\rm peak}=1500$\,keV, $\alpha=-1.5$.  For the preferred model\footnote{For the ``hard'' spectral model the limit would be shallower by a factor of $\sim 2$, although given the $E_{\rm peak}-E_{\rm iso}$ relation \cite{Amati2006} a spectrally hard burst at $z=1.29$ would be expected to also be very luminous.}, the typical limit on the 1\,s peak flux 
during the optical explosion time window is
$F < 9\times10^{-8}$\,erg\,cm$^{-2}$\,s$^{-1}$,
equivalent to a limit on the peak luminosity of 
$L_{\rm iso,peak} < 8\times10^{50}$\,erg\,s$^{-1}$.
For 8.192\,s intervals, the limit on the average flux is $<3\times10^{-8}$\,erg\,cm$^{-2}$\,s$^{-1}$, equivalent to $L_{\rm iso,peak} < 2.8\times10^{50}$\,erg\,s$^{-1}$.

To convert these values to approximate limits on the burst fluence, we take the 8\,s flux limit and multiply by the assumed characteristic (observed) timescale, typically 40\,s for long-duration GRBs.  We obtain $S < 1.2\times10^{-6}$\,erg\,cm$^{-2}$, equivalent to $E_{\rm iso} < 5\times10^{51}$\,erg.

{\it Fermi} was occulted in the direction of AT\,2019pim for about 20 min at the beginning of the afterglow-inferred explosion window and about 40 min toward the end of the window, so no limit can be placed on gamma-ray emission during this period from GBM.  However, the Interplanetary Network was sensitive to the position of AT\,2019pim throughout this interval, and no detections are recorded.

From the Konus-Wind observations, using the same spectral model as for the GBM upper limit calculations, the 90\% confidence limiting peak flux (10--1000\,keV, 2.944~s timescale) is $1.5 \times 
10^{-7}$\,erg\,cm$^{-2}$\,s$^{-1}$, equivalent to $L_\mathrm{iso,peak} < 4.4 \times 10^{51}$\,erg\,s$^{-1}$. Assuming a similar scaling over longer intervals as in GBM, the equivalent $E_\mathrm{iso}$ limit is about $E_\mathrm{iso} < 3.6 \times 10^{52}$\,erg.

The position of AT\,2019pim was not\footnote{Search conducted using \url{https://github.com/lanl/swiftbat\_python}} in the field of view of the Swift Burst Alert Telescope (BAT; \citealt{Barthelmy2004}), except for during short windows.

\subsection{Swift XRT Observations}

We obtained two 3\,ks observations with the
X-Ray Telescope (XRT; \citealt{Burrows2005}) on board
the {\it Neil Gehrels Swift observatory} ({\it Swift}; \citealt{Gehrels2004})
under a target-of-opportunity program (target ID 11549).
The first observation started on Sept 4.13,
and the second started on Sept 12.08.
Using the online tool\footnote{\url{https://www.swift.ac.uk/user_objects/}} from the {\it Swift} team \citep{Evans2007,Evans2009},
we found that the count rate in the first observation was
$0.021 \pm 0.003\,$s$^{-1}$
with a best-fit photon index of $\Gamma=1.8^{+0.8}_{-0.6}$
and a corresponding unabsorbed flux density of $f_X = 8.6^{+4.6}_{-2.5}\times10^{-13}\,$erg\,cm$^{-2}$\,s$^{-1}$ (90\% confidence),
and $L_X = 2.5^{+1.4}_{-0.7} \times 10^{46}$\,erg\,s$^{-1}$ \citep{GCN25658}.
This assumes a neutral hydrogen column density
$n_H = 5.6 \times 10^{20}$\,cm$^{-2}$ \citep{Willingale2013}.
In the second observation, the count rate was
$0.003 \pm 0.001\,$s$^{-1}$.
Assuming the same photon index ($\Gamma=1.8$)
and $n_H$ we used webpimms\footnote{\url{https://heasarc.gsfc.nasa.gov/cgi-bin/Tools/w3pimms/w3pimms.pl}} to find
$f_X = (1.3 \pm 0.5)\times10^{-13}\,$erg\,cm$^{-2}$\,s$^{-1}$
and
$L_X = (3.8 \pm 1.5) \times 10^{45}$erg\,s$^{-1}$.

\subsection{Radio Observations}

Shortly after the spectroscopic confirmation of the transient, we triggered our pre-approved VLA program for follow-up observations of orphan afterglows (program ID VLA/18B-242, PI D. Perley).  The transient was well detected in the initial X-band observation and we continued following it with a series of observations at L, S, C, X, and Ku bands during the 2019B A-configuration cycle.  Observations in different bands were not always obtained at the same epoch owing to scheduling constraints.  Late-time observations were obtained in 2020 via dedicated follow-up programs (IDs VLA/19B-342 and VLA/20A-506, PI D. Perley).  This included a D-configuration observation in C, X, and Ku bands in January 2020, a C-configuration observation in X and Ku bands in April 2020, and a final deep ($t_{\rm int}=2.15$\,hr) C-configuration observation in X band in June 2020.   The D-configuration C-band observations were significantly affected by radio frequency interference (RFI), as were the April C-configuration X-band observations above 10\,GHz.

Data reduction was performed using standard procedures in the Astronomical Image Processing System (AIPS).  Images were made in separate windows with a bandwidth of 1\,GHz, except in the last two observations where images were made with a 2\,GHz bandwidth.  Flux-density measurements were performed using \texttt{jmfit}.  In the small number of cases where the afterglow was not securely detected, the location of the centroid was fixed to the position as measured in our high-S/N A-configuration imaging to provide a forced measurement of the flux density.  All values are reported in Table \ref{tab:vla}.  Reported uncertainties do not include errors in the flux calibration, which is expected to be about 5\% (or less) of each measurement.

We do not apply any corrections for radio emission from the host galaxy.  The star-formation rate of the host as measured from optical SED fitting (\S \ref{sec:hostsedfit}) is about 3~$M_\odot$\,yr$^{-1}$, which (using the relations in \citealt{Murphy+2011}) at the distance of AT2019pim would contribute only $\sim 0.6$\,$\mu$Jy of radio continuum flux at 1 GHz and less at higher frequencies, and so can safely be ignored.

\begin{table*}
	\centering
	\caption{VLA measurements of AT\,2019pim}
	\label{tab:vla}
	\begin{tabular}{lrrr|@{\hskip 1in}lrrr}
		\hline
MJD & $\nu$ & $F_\nu$ & unc. & MJD & $\nu$ & $F_\nu$ & unc. \\
    &  (GHz) & ($\mu$Jy)  & ($\mu$Jy) &     &  (GHz) & ($\mu$Jy)  & ($\mu$Jy) \\
		\hline
58731.0379 &  8.50 &   25 &   9 & 58760.1055 &  5.50 &  340 &   9 \\
58731.0379 &  9.50 &   52 &   8 & 58760.1055 &  6.50 &  451 &   9 \\
58731.0379 & 10.50 &   93 &   9 & 58760.1055 &  7.50 &  434 &   9 \\
58731.0379 & 11.50 &  103 &  10 & 58760.1278 &  8.50 &  345 &   9 \\
58733.1521 &  8.50 &  183 &   9 & 58760.1278 &  9.50 &  314 &   9 \\
58733.1521 &  9.50 &  158 &   9 & 58760.1278 & 10.50 &  298 &   9 \\
58733.1521 & 10.50 &  139 &  15 & 58760.1278 & 11.50 &  293 &  10 \\
58733.1521 & 11.50 &  178 &  15 & 58760.1535 & 12.50 &  253 &  10 \\
58737.1719 &  8.50 &  237 &   9 & 58760.1535 & 13.50 &  246 &   9 \\
58737.1719 &  9.50 &  233 &   8 & 58760.1535 & 14.50 &  257 &  10 \\
58737.1719 & 10.50 &  239 &  12 & 58760.1535 & 15.50 &  253 &  10 \\
58737.1719 & 11.50 &  248 &  12 & 58760.1535 & 16.50 &  243 &  11 \\
58739.1205 &  2.25 &    0 &  30 & 58760.1535 & 17.50 &  266 &  13 \\
58739.1205 &  2.75 &  154 &  16 & 58770.0915 &  1.02 &   29 &  54 \\
58739.1205 &  3.25 &  163 &  12 & 58770.0915 &  1.28 &  193 &  35 \\
58739.1205 &  3.75 &  143 &  12 & 58770.0915 &  1.52 &  188 &  46 \\
58739.1430 &  8.50 &  146 &   9 & 58770.0915 &  1.78 &  245 &  38 \\
58739.1430 &  9.50 &  157 &   8 & 58770.1140 &  2.25 &  131 &  19 \\
58739.1430 & 10.50 &  178 &   9 & 58770.1140 &  2.75 &  141 &  15 \\
58739.1430 & 11.50 &  212 &  10 & 58770.1140 &  3.25 &  119 &  12 \\
58739.1684 & 12.50 &  265 &  10 & 58770.1140 &  3.75 &  108 &  12 \\
58739.1684 & 13.50 &  243 &   9 & 58775.0573 &  8.50 &  161 &   7 \\
58739.1684 & 14.50 &  280 &  10 & 58775.0573 &  9.50 &  169 &   7 \\
58739.1684 & 15.50 &  265 &  10 & 58775.0573 & 10.50 &  166 &   7 \\
58739.1684 & 16.50 &  315 &  11 & 58775.0573 & 11.50 &  129 &   8 \\
58739.1684 & 17.50 &  306 &  12 & 58775.9458 &  4.50 &  166 &  10 \\
58739.1927 &  5.00 &  153 &   7 & 58775.9458 &  5.50 &  114 &   9 \\
58739.1927 &  7.00 &  139 &   6 & 58775.9458 &  6.50 &  147 &   9 \\
58739.1927 &  4.50 &  142 &  10 & 58775.9458 &  7.50 &  118 &   8 \\
58739.1927 &  5.50 &  138 &   9 & 58775.9677 &  8.50 &  142 &   8 \\
58739.1927 &  6.50 &  129 &   9 & 58775.9677 &  9.50 &  121 &   8 \\
58739.1927 &  7.50 &  144 &   8 & 58775.9677 & 10.50 &  146 &   8 \\
58745.0569 &  2.75 &  194 &  16 & 58775.9677 & 11.50 &  162 &   9 \\
58745.0569 &  3.25 &  254 &  12 & 58775.9934 & 12.50 &  143 &   9 \\
58745.0569 &  3.75 &  429 &  11 & 58775.9934 & 13.50 &  153 &   8 \\
58745.0792 &  8.50 &  323 &   9 & 58775.9934 & 14.50 &  127 &   8 \\
58745.0792 &  9.50 &  342 &   9 & 58775.9934 & 15.50 &  149 &   8 \\
58745.0792 & 10.50 &  366 &   9 & 58775.9934 & 16.50 &  119 &   9 \\
58745.0792 & 11.50 &  357 &  10 & 58775.9934 & 17.50 &  134 &  10 \\
58745.1048 & 12.50 &  378 &   9 & 58860.9094 &  9.00 &   41 &   7 \\
58745.1048 & 13.50 &  405 &   9 & 58860.9094 & 11.00 &   34 &   7 \\
58745.1048 & 14.50 &  396 &   9 & 58860.9441 & 12.50 &   49 &   8 \\
58745.1048 & 15.50 &  399 &   9 & 58860.9441 & 13.50 &   40 &   7 \\
58745.1048 & 16.50 &  419 &  10 & 58860.9441 & 14.50 &   27 &   7 \\
58745.1048 & 17.50 &  420 &  12 & 58860.9441 & 15.50 &   43 &   8 \\
58745.1288 &  4.50 &  408 &   9 & 58860.9441 & 16.50 &   34 &   8 \\
58745.1288 &  5.50 &  379 &   9 & 58860.9441 & 17.50 &   47 &   8 \\
58745.1288 &  6.50 &  324 &   8 & 58860.9760 &  4.50 &   55 &  19 \\
58745.1288 &  7.50 &  329 &   8 & 58860.9760 &  5.50 &   88 &  25 \\
58757.0172 &  1.02 &   96 &  50 & 58860.9760 &  6.50 &   88 &  17 \\
58757.0172 &  1.28 &  121 &  36 & 58860.9760 &  7.50 &   33 &  19 \\
58757.0172 &  1.52 &  108 &  49 & 58940.6566 &  9.00 &   23 &   8 \\
58757.0172 &  1.78 &  -19 &  42 & 58940.6997 & 12.77 &   20 &   7 \\
58757.0397 &  2.25 &  159 &  47 & 58940.6997 & 14.30 &   16 &   6 \\
58757.0397 &  2.75 &   74 &  19 & 58940.6997 & 15.84 &   20 &   6 \\
58757.0397 &  3.25 &   92 &  17 & 58940.6997 & 17.38 &   27 &   8 \\
58757.0397 &  3.75 &  103 &  13 & 59004.5502 &  9.00 &   14 &   2 \\
58760.1055 &  4.50 &  190 &  10 & 59004.5502 & 11.00 &   10 &   3 \\
		\hline
	\end{tabular}
\end{table*}

\section{Empirical Modeling and Analysis}
\label{sec:modeling}

Before interpreting the emission physically, we first attempt to fit simple empirical models to constrain key features: specifically the explosion time, temporal and spectral slopes, and temporal and spectral breaks.

\subsection{Explosion Time and Early Decay}
\label{sec:explosiontime}

We fit an empirical model to all ground-based optical photometry within 4\,days after discovery (plus all TESS data in the range MJD 58727.657--58729.158, or approximately $-0.5$\,d to +1.0\,d after the ZTF discovery observation).  We initially assume a simple broken \cite{Beuermann+1999} power law with the onset time and peak time being free parameters, although we later extend this to add a second additive power law and a jet break at later times (\S \ref{sec:lateoptical}).  The sharpness parameter was fixed at 0.5, and the power-law index of the rising phase of the afterglow is fixed to $\alpha_{\rm rise} = +3.0$ (as expected for optical/X-ray emission from a relativistic, constant velocity thin shell expanding into a uniform medium).
The evolution of the afterglow is assumed to be achromatic but the relative flux in each band is a free parameter.

\begin{figure}
\includegraphics[width=0.47\textwidth]{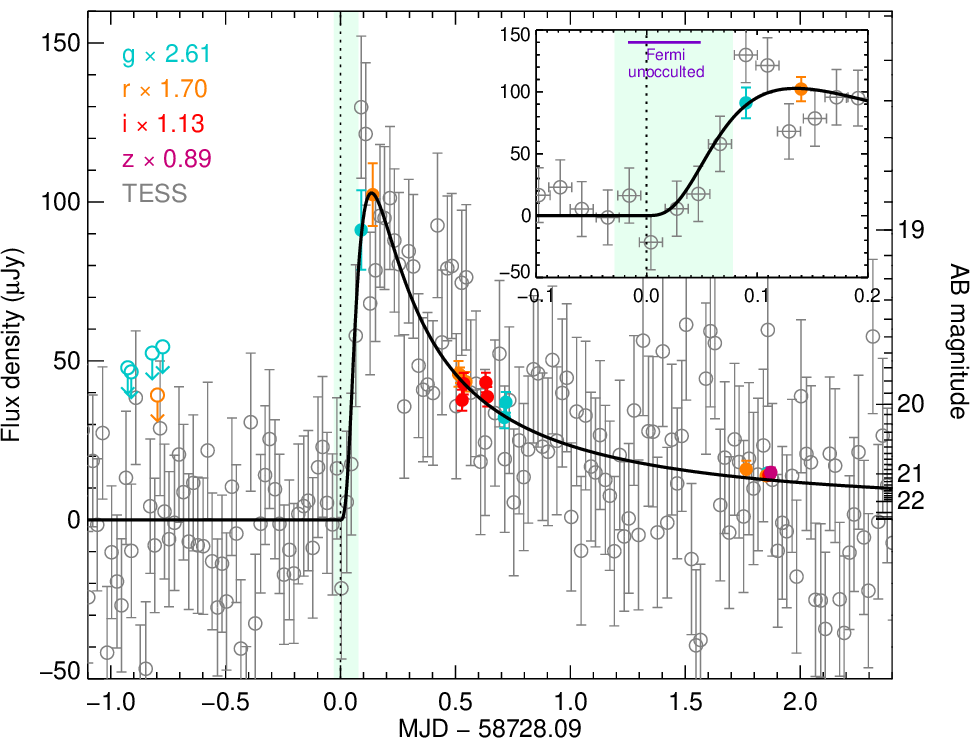}
\caption{Early-time observations of AT\,2019pim from TESS, P48, GIT, and LT.
The black curve shows a broken power-law model fit to the TESS and ground-based data simultaneously; the flux scale for the ground-based filters has been shifted to align the data using this model.  The shaded region shows a conservative bracketing of the potential explosion time, with the best-fit $t_0$ (for an assumed $\alpha_{\rm rise}=3.0$ and $s=0.5$) indicated with a dotted vertical line.  An inset focusing in on the region around the explosion time is shown at top right; the {\it Fermi}-GBM sensitivity window (\S \ref{sec:grblimits}) is shown.}
\label{fig:tess}
\end{figure}

\begin{figure}
\includegraphics[width=0.47\textwidth]{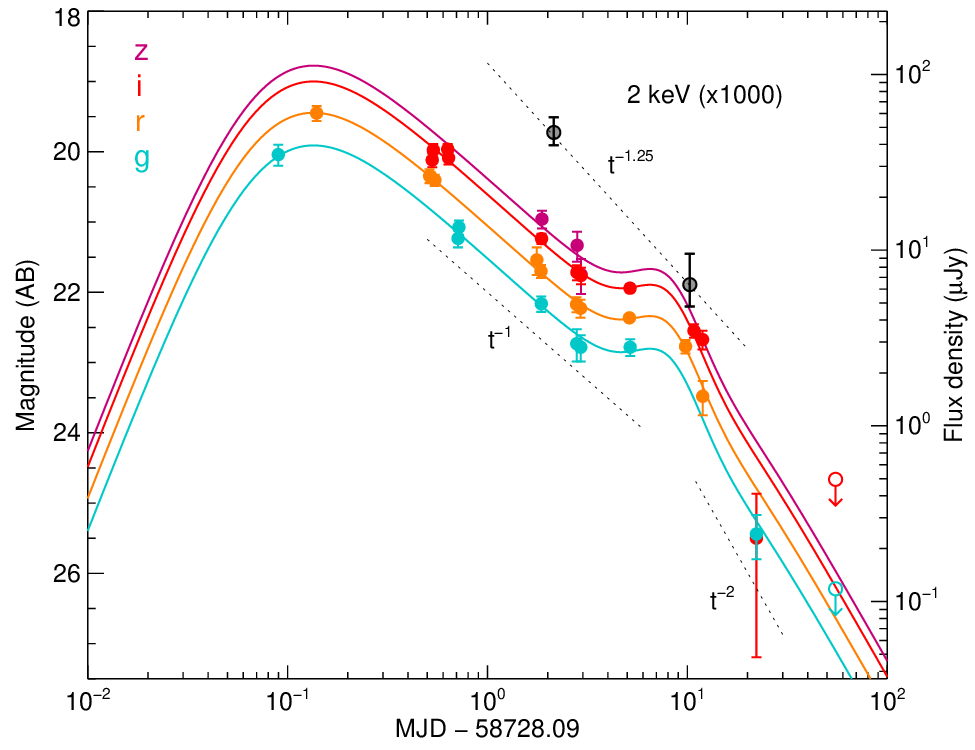}
\caption{Complete ground-based optical light curve of AT\,2019pim, on a logarithmic scale; the two XRT detections are also shown (rescaled by a factor of $10^3$).  The time axis is relative to the preferred explosion time of our model, although we emphasise that this is uncertain.  The light curve exhibits an approximately $t^{-1}$ decay before briefly leveling off, then rapidly steepening.}
\label{fig:lc}
\end{figure}

The resulting best-fit curve is plotted in Figure~\ref{fig:tess} and in Figure~\ref{fig:lc}.  The model indicates a peak close in time to the first ZTF detection and an explosion time $\sim 2$\,hr prior: for our assumed $\alpha_{\rm rise} = +3.0$ and $s=0.5$ we obtain an explosion time (MJD) of 58728.0898 $\pm$ 0.0289 (2$\sigma$) although this is strongly sensitive to those assumptions, and for a more sudden initial rise the explosion time can be significantly more recent.   The TESS measurement centred at MJD 58728.1573 is 2.6$\sigma$ above the background, so we have reasonable confidence that the afterglow began to rise sometime within or before this exposure (i.e., no later than MJD 58728.168), placing a firm upper limit on the explosion time.  A conservative bracketing of the exposure time combining these constraints is shown as the shaded region in Figure~\ref{fig:tess} (MJD 58728.062--58727.168).

\subsection{Late Plateau and Break}
\label{sec:lateoptical}

A single-component power-law fit to the light curve over the first five days suggests a post-peak decay index of  $\alpha \approx -1$.  However, the decay behaviour is clearly more complicated than this:  between about 5 to 10 days the rate of decay briefly becomes much shallower, before then steepening dramatically, and there are no detections of the afterglow beyond 20 days even in deep Keck imaging.  

To incorporate this behaviour, we introduced a second \cite{Beuermann+1999} broken power-law component to the model described in \S\,\ref{sec:explosiontime} (which adds in flux space to the initial component; see e.g. Equation 1 of \citealt{Perley+2008}) as well as a late-time break at 20 days to an assumed final decay index of $\alpha_{\rm late} = -2$.  While this model is not unique (owing to the sparse nature of the post-plateau follow-up observations, it is not possible to robustly fit all parameters), it provides a good match to all the data and is used consistently to visualize the early-through-late-time optical light curve in subsequent figures.

\subsection{Radio Light Curve}

\begin{figure}
\includegraphics[width=0.47\textwidth]{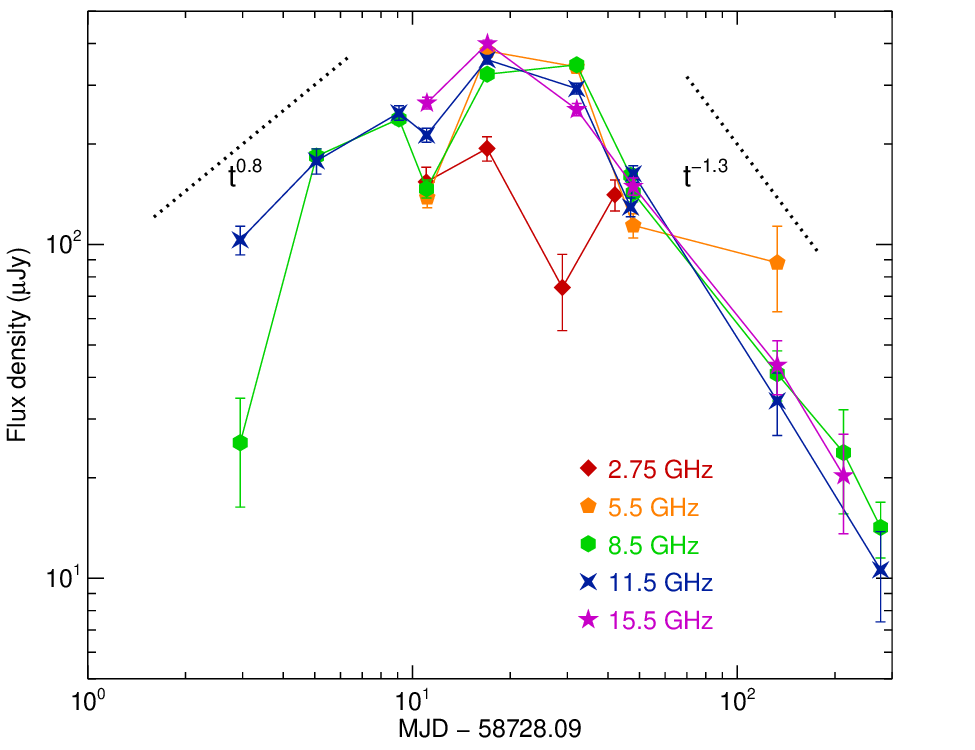}
\caption{VLA radio light curve of AT\,2019pim at a few select frequencies.  The high-frequency ($>$\,10\,GHz) light curve shows relatively consistent behaviour, with a gradual rise followed by a decline.  Lower frequencies exhibit strong interepoch variability out to late times, likely due to interstellar scintillation.}
\label{fig:radio}
\end{figure}

The radio light curve at a few select frequencies is shown in Figure~\ref{fig:radio}.  It exhibits a gradual rise ($F \propto t^{+0.8}$ at high frequencies), peaks $\sim 20$\,d post-discovery, and then fades ($t^{-1.3}$).  Significant short-timescale variability is superimposed on top of this slow evolution, especially at the lower frequencies ($<10$\,GHz) and early times ($t<50$\,days).

\begin{figure}
\includegraphics[width=0.47\textwidth]{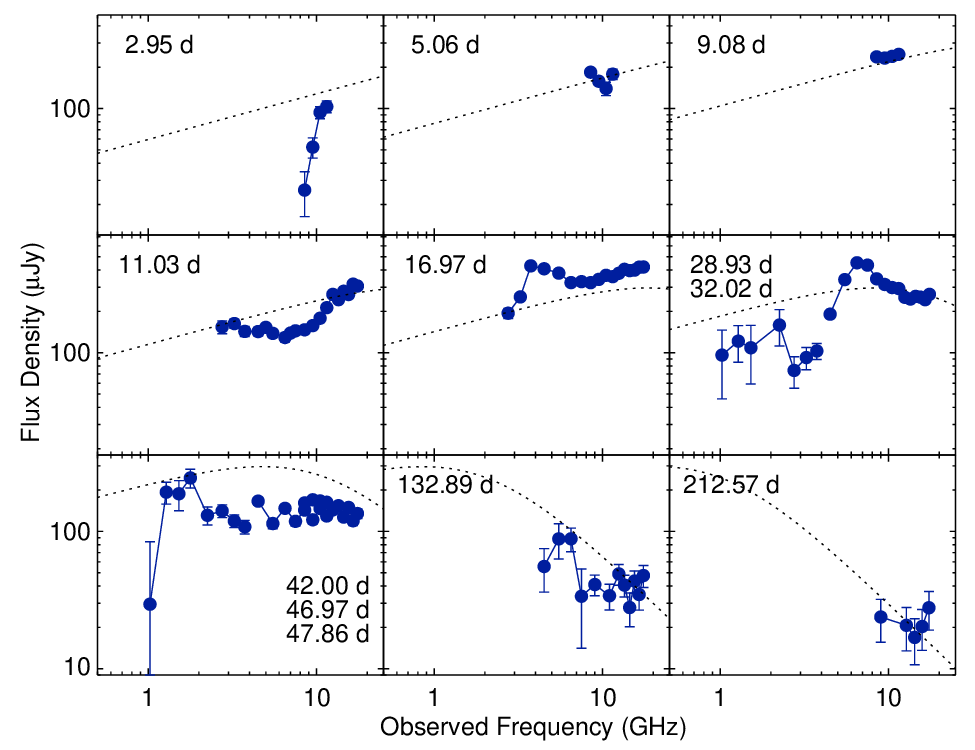}
\caption{Radio SEDs of AT\,2019pim at different epochs (the observer-frame days elapsed since the preferred explosion time are indicated in each panel).  In two cases the low- and high-frequency blocks were observed on separate days.  A simple model of the afterglow SED assuming standard ISM is shown.  While the spectra broadly (within a factor of $\sim 2$) follow this model, the deviations over narrower bandwidths are significant, probably due to interstellar scintillation.
}
\label{fig:radioseds}
\end{figure}

The rapid low-frequency variability results in complex radio spectra.  Approximately coeval SEDs are shown in Figure \ref{fig:radioseds}.  Between 10--30\,days the SEDs cannot be well-fit with a power law (or broken power law) owing to modulations in the SED by a factor of $\sim 1.5$--2, producing structure on a frequency scale of $\Delta\nu/\nu \approx 2$.  This behaviour is present until at least 30\,days, and may persist beyond that (although the more limited frequency coverage and lower S/N makes it difficult to be definitive).  The average spectral index ($F_\nu \propto  \nu^{\beta}$) (forcing a power-law fit to each spectrum for which multiple receivers were used, excluding measurements below 3\,GHz) is typically about $\beta=+0.1$, although it ranges between $-0.5$ and +0.6.

This single-peaked behaviour is typical of GRB light curves, as the spectral break associated with the minimum synchrotron energy $\nu_m$ passes through the radio bands.  It is difficult to clearly identify this break in any of the available radio spectra (shown in Figure~\ref{fig:radioseds}) as a result of what is likely quite strong interstellar scintillation (\S \ref{sec:scintillation}). 
However, the well-sampled multiband SEDs at $\Delta t \approx 11$\,d and $\Delta t \approx 17$\,d are broadly consistent with the $F_\nu \propto \nu^{+1/3}$ spectrum expected below the synchrotron peak (suggesting $\nu_m > 10$\,GHz at this time), the $\Delta t \approx 45$\,d spectrum is largely flat (suggesting $\nu_m \approx 10$\,GHz), and the $\Delta t \approx 132$\,d spectrum, while having low S/N, shows a negative spectral index (suggesting $\nu_m < 10$\,GHz); this is broadly consistent with the expected passage of $\nu_m$ through the radio band for a relativistically expanding outflow.  However, the strong scintillation and lack of low-frequency coverage during the D/C-configuration cycles do not allow us to robustly model the behaviour in more detail, or to easily discriminate between constant-density or $r^{-2}$ density profiles.  There is no obvious counterpart of the ``bump'' and corresponding sharp dropoff seen in the optical light curve at 10 days, although the peak of the radio light curve occurs only a few days after this.

\subsection{Optical SED and Extinction Column}

The optical transient is quite red.  We extracted the simultaneous $griz$ SED of the transient using the LT data 
 1--3\,days post-explosion and applied a Galactic extinction correction ($E_{B-V}=0.038$\,mag; \citealt{Schlafly+2011}).
A power-law fit to these data implies an apparent spectral index of $\beta = -1.70 \pm 0.14$ (using the convention $f_\nu \propto \nu^\beta$).  The optical to X-ray spectral index at the same time is significantly shallower ($\beta_{OX}=-0.92$), implying that the optical flux is likely extinguished by moderate host-galaxy dust.  The blue portion of the LRIS spectrum also shows slight curvature at approximately the expected location of the redshifted 2175\,\AA\ extinction feature commonly seen in local galaxies.

To constrain the extinction column, we assume an intrinsic optical spectral index of $\beta=-0.65$ (\S \ref{sec:modeling2}) and
adopt a \cite{Fitzpatrick1999} dust-extinction law (see also \citealt{Fitzpatrick+1988,CCM1989,Fitzpatrick+1990}) with the values of most of the parameters set to their diffuse Milky Way values, with the exception that the strength of the 2175\,\AA\ bump is allowed to be a free parameter ($c3$).  We find a good fit to our spectrum for $E_{B-V} = 0.24$\,mag and $c3=1.0$ (red curve in Figure \ref{fig:spectrum}), implying host extinction of about 1\,mag in the observed optical bands.

\subsection{Host-Galaxy SED}
\label{sec:hostsedfit}

\begin{figure}
\includegraphics[width=0.48\textwidth]{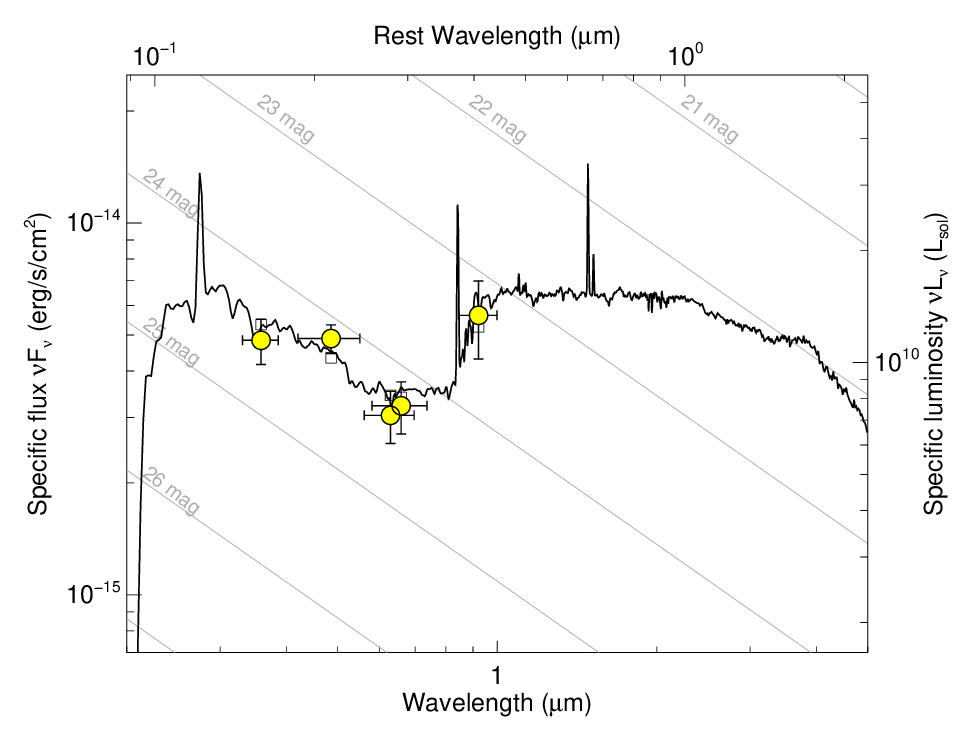}
\caption{Population-synthesis fit to the host galaxy SED.  Yellow-filled circles indicate host photometry; empty squares show synthetic photometry of the afterglow over these bands.  Measurements are corrected for Galactic extinction.  The best-fit model is for a star-formation rate of 2.7\,$M_\odot$\,yr$^{-1}$ and stellar mass of $M_* = 1.9 \times 10^{10}\,M_\odot$, typical of GRB host galaxies at these redshifts.}
\label{fig:hostsedfit}
\end{figure}

The late-time filter coverage is (marginally) sufficient to obtain basic constraints on the fundamental properties of the host galaxy using SED fitting.  We use codes previously employed by \cite{Perley+2014} and population-synthesis templates from \cite{BC03} to fit the $ugrRz$ data against a model that assumes a single stellar population with a uniform star-formation history and \cite{Calzetti+1994} dust attenuation.  The data are well fit by a model with a moderate star-formation rate (SFR = $2.7^{+4.0}_{-1.1}$\,$M_\odot$\,yr$^{-1}$), moderate stellar mass ($M_* = 1.9^{+0.5}_{-1.5} \times 10^{10}\,M_\odot$), and low to moderate dust extinction ($A_V = 0.18^{+0.36}_{-0.18}$\,mag).  These properties are typical of star-forming galaxies (and of long-GRB hosts) at similar redshifts.  A plot of the SED is given in Figure \ref{fig:hostsedfit}.

\section{Discussion}

\subsection{AT\,2019pim in Context: Empirical Constraints on a GRB Origin}

As the first afterglow with both a confirmed redshift and strong limits on an associated gamma-ray counterpart, AT\,2019pim is of interest primarily as a candidate for a phenomenon related to but separate from ``typical'' GRBs:  a dirty fireball, an off-axis GRB, or a GRB with a smooth outflow free of internal shocks.  

Before considering these possibilities, it should be emphasised that even the known GRB phenomenon (as selected by existing high-energy satellites) is extremely diverse:  there are numerous examples of ``normal'' but lower-luminosity GRBs\footnote{We distinguish this population (with $E_{\rm iso} = 10^{50}-10^{51}$\,erg) from the truly low-luminosity GRBs such as GRB\,980425 or GRB\,060218 whose inferred energy outputs are orders of magnitude lower ($E_{\rm iso} = 10^{48}-10^{50}$\,erg) and which could in principle be a separate population \citep{Liang+2007,Virgili+2009,Bromberg+2011,Nakar2015}.} which are visible in the low-redshift universe yet would not be detectable to {\it Konus}, {\it Fermi}, or even {\it Swift} at higher redshifts (e.g., \citealt{Singer+2013,Schulze+2014,Dichara+2022}).
Thus, as a first step, it is important to establish that AT\,2019pim stands out from the well-established GRB and afterglow population in at least some way.

We restrict our comparisons to long-duration GRBs specifically.  While short GRBs may also produce afterglows, their optical luminosities are typically much lower, and even among the existing gamma-ray-selected population few or none would be detectable by ZTF at the distance of AT\,2019pim \citep{Kann+2011}.  Certain tidal disruption events also produce relativistic ``afterglows'' \citep{Bloom+2011,Levan+2011,Zauderer+2011,Burrows+2011,Andreoni+2022}, but these have quite distinctive X-ray and radio behaviour different from this event.  The association of this event with a star-forming low-mass galaxy (\S \ref{sec:hostsedfit}) further supports this.

\begin{figure*}
\includegraphics[width=0.97\textwidth]{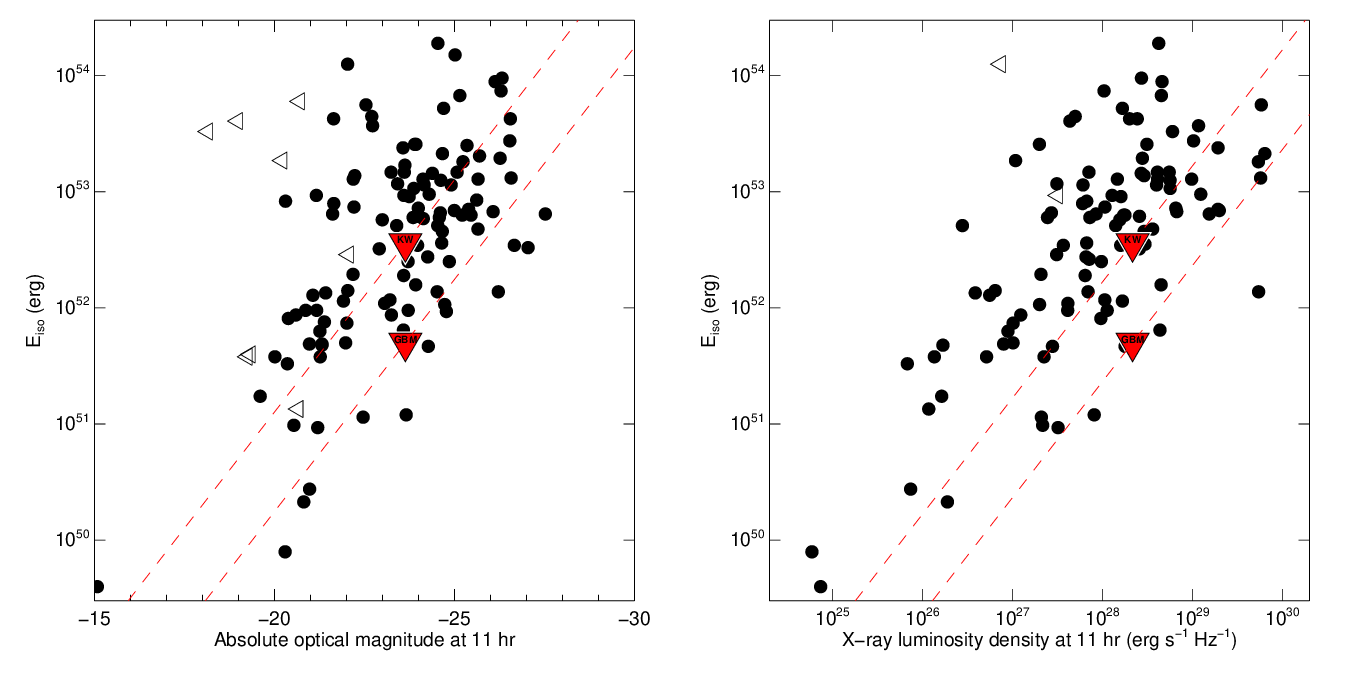}
\caption{Afterglow luminosity versus 10-1000 keV prompt emission isotropic-equivalent energy for GRBs in the optical (left panel) and X-rays (right panel).   Upper limits on the prompt emission for AT\,2019pim are shown as red triangles:  the lower filled triangle is the GBM limit (for an event close in time to our best-fit explosion time) and the upper triangle is the {\it Konus} limit (a more conservative limit allowing a burst during the GBM occulations).  Most of the known GRB population for afterglows of comparable luminosities is ruled out by the GBM limit, although not by the {\it Konus} limit.}
\label{fig:promptafterglow}
\end{figure*}

Figure \ref{fig:promptafterglow} shows the afterglow luminosity at the commonly standardised time of 11\,hr post-GRB (observed) versus the prompt emission $E_{\rm iso,\gamma}$ for a sample of pre-{\it Swift} and early-{\it Swift} bursts (from \citealt{Nysewander+2009}).  GRB fluences were converted from the 15-150 keV band to the 10-1000 keV band using an average correction factor of 2.39, derived from our preferred spectral model.  Afterglow luminosities are calculated assuming a basic $K$-correction factor of $1+z$.  Luminosities could be further corrected to standard times and frequencies in the \emph{rest} frame as $F_{\rm rest} = F_{\rm obs}(1+z)^{\alpha}/(1+z)^{\beta}$, assuming a GRB light curve power-law index of $\alpha$ and spectral index of $\beta$; for typical afterglows at these frequencies and timescales $\alpha \approx \beta$ $(\approx -1)$, these factors helpfully cancel out, and we neglect this correction.  Under these assumptions, the left plot can be treated as an $R$-band absolute magnitude at 11 rest-frame hours or equivalently as a 3000\,\AA\ absolute magnitude at 5 rest-frame hours; the right plot can be treated as a 1\,keV rest-frame luminosity density at 11 rest-frame hours or a 2.2\,keV rest-frame luminosity density at 5 rest-frame hours.

The GBM limit on GRB emission from AT\,2019pim is shown as the lower of the two solid red triangles in each panel of Figure \ref{fig:promptafterglow}.   The X-ray flux is extrapolated backward to 11\,hr assuming $\alpha = -1$.  
The bulk of known GRBs with comparable afterglow luminosities have prompt emission substantially brighter (by a factor of 10--30) than what the {\it Fermi} limit allows for AT\,2019pim. 
Thus, assuming that the explosion did indeed occur at or close to our inferred explosion time, this event is clearly uncharacteristic of the ``normal'' GRB population (if not completely unprecedented: a handful of GRBs with lower $E_{\rm iso}$ values do have comparable afterglows).

The GBM limit covers the most probable time of explosion but (due to occultations) does not cover the entire allowed explosion time window.  The shallower limit from {\it Konus} is also shown as the upper triangle in Figure \ref{fig:promptafterglow}.  This also rules out most GRBs of comparable afterglow luminosity, but a substantial fraction of the population does lie below the {\it Konus} limit, and so we cannot fully rule out a GRB scenario from high-energy limits alone.  However, the early light curve would be unusual for a GRB occurring during either of the occultations:  a GRB in the first occultation would have an unusually long rise time of almost 3\,hr; a GRB in the second occultation would have to exhibit a fast rise time and then a multi-hour plateau with virtually no fading.

\begin{figure}
\includegraphics[width=0.47\textwidth]{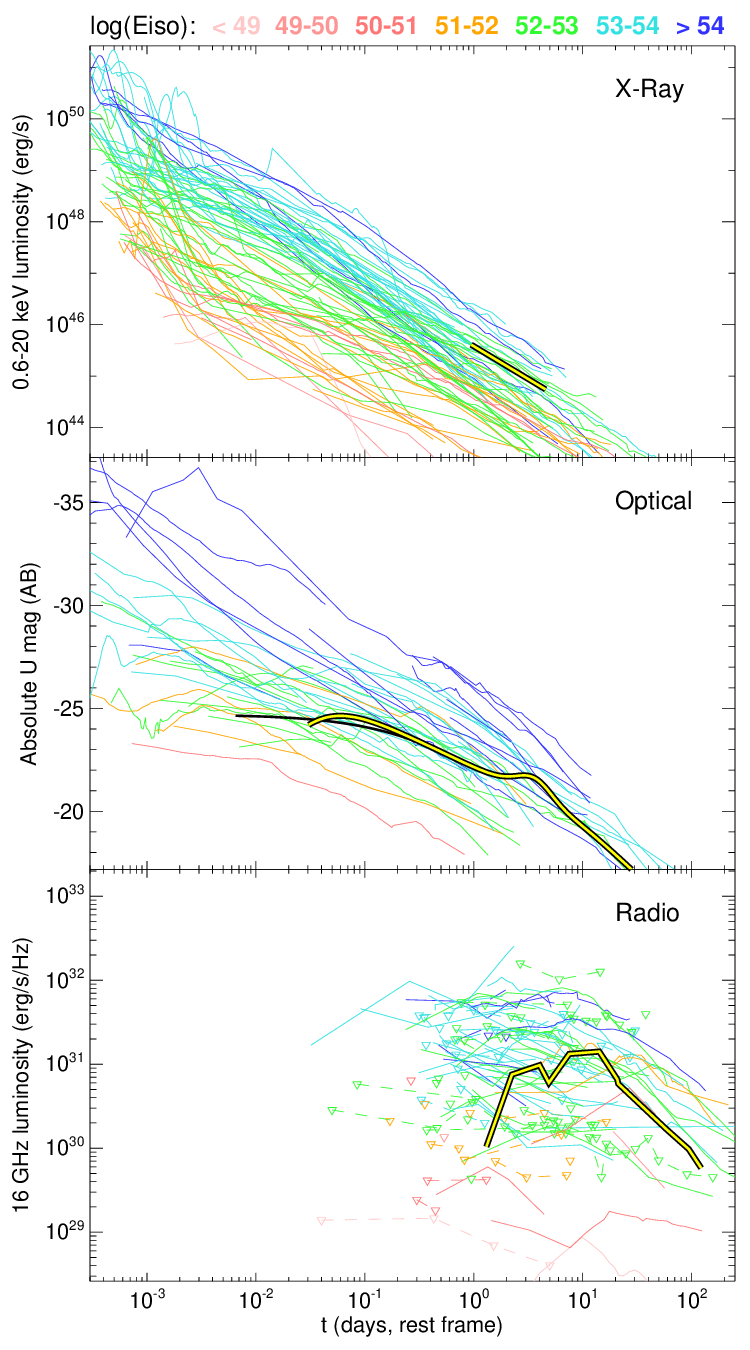}
\caption{Comparison between the X-ray, optical, and radio light curves of AT\,2019pim (thick yellow/black line) and GRB afterglows (colour-coded by $E_{\rm iso}$).  In the optical panel, the yellow/black curve represents our empirical model, starting from the time of the first TESS detection; the solid-black line indicates an alternative fit with the explosion time $t_0$ set to the end of our bracketed window.   The afterglow is similar in luminosity (in X-ray, optical, and radio bands) to GRB afterglows with $E_{\rm iso} \approx 10^{52}-10^{53}$\,erg.}
\label{fig:compareplot}
\end{figure}

We can also perform comparisons of this type more qualitatively over the entire light curve to compare the general behaviour and time-dependent luminosity to the general afterglow population.  Figure \ref{fig:compareplot} shows the X-ray, optical, and radio light curves of this event in comparison to GRBs, colour-coded by $E_{\rm iso}$.  The comparison population is the same as in the equivalent figure of \cite{Perley+2014}: specifically, events from the sample analyses of \cite{Evans2007}, \cite{Cenko+2009}, \cite{Kann+2011}, and \cite{Chandra+2012}.  The luminosity and general decay rate of AT\,2019pim are fairly typical at \emph{late} times, although at every wavelength the luminosity is characteristic only of high-$E_{\rm iso}$ ($10^{52}-10^{54}$\,erg) events, as expected given Figure \ref{fig:promptafterglow}.  Comparing the early phase is more difficult owing to the uncertain explosion time of AT\,2019pim, but rise times as slow as 0.5 rest-frame hours are rare, constituting no more than a few percent of known afterglows with early-time follow-up observations (although a few examples do exist, e.g., \citealt{Margutti+2010}).  This has also been noted in other early-afterglow samples \citep{Rykoff+2009,Melandri+2014a,Hascoet+2014,Ghirlanda+2018,Oates+2019,Jayaraman+2023}.

\section{Physical Constraints on the Outflow}

\subsection{Constraints from Basic Physical Arguments}

Even in the absence of a complete model to explain the multiwavelength behaviour of the afterglow across all bands, the observations can be used directly to constrain the source size and therefore outflow velocity.  Three independent constraints are possible: a lower limit from the emergence (rise) time of the afterglow, a lower limit from the nonthermal spectrum, and an upper limit based on the presence of strong scintillation at late times.

\subsubsection{Constraint from rise time of afterglow}
\label{sec:risetimelimit}

The afterglow forward shock reaches peak luminosity when the ejecta have had time to sweep up sufficient material from the surrounding medium to gain mass-energy approximately comparable to that of the initial outflow (for a review see, e.g., \citealt{Meszaros2006}).   Time-of-flight effects greatly compress this characteristic timescale for material moving toward the observer at relativistic speeds, making the early afterglow a particularly sensitive probe of the Lorentz factor.  In the case of a uniform and wind-driven medium, respectively, 
the equations relating the observed deceleration time $t$ to the maximum Lorentz factor $\Gamma$ are

$$\Gamma = k_{0} \left(\frac{E_K}{n m_p c^5}\right)^{1/8} \left(\frac{t}{1+z}\right)^{-3/8}\, ,$$
$$\Gamma = k_{2} \left(\frac{E_K}{A m_p c^3}\right)^{1/4} \left(\frac{t}{1+z}\right)^{-1/4}\, .$$

\noindent Here $E_K$ is the isotropic-equivalent energy of the outflow, $n$ is the density of the circumburst interstellar medium (ISM),
and $A$ is the wind density parameter ($\rho = A r^{-2}$).  
The numerical prefactors $k_0$ and $k_2$ have values of order unity but
vary slightly according to different authors (we adopt $k_0=0.65$ and $k_2=0.45$, following \citealt{Sari+1999}; see \citealt{Ghirlanda+2018} for a compilation of alternative values).

The TESS observations strongly suggest an afterglow rise time of 1--4\,hr (observer frame).  If this rise is the result of deceleration of the afterglow, the corresponding fiducial ranges of the Lorentz factor
in the uniform and wind cases, respectively, are

$$28 \lesssim \frac{\Gamma}{(E_{53}/n_0)^{1/8}} \lesssim 47\, , $$
$$9 \lesssim \frac{\Gamma}{(E_{53}/A*)^{1/4}} \lesssim 14\, .$$

\noindent Here $E_{53} = E_K / (10^{53}\,{\rm erg})$, $n_0 = n / {\rm cm}^{-3}$, %.)}
and $A* = A/(3\times10^{35} {\rm\ g\,cm^{-1}})$.  The lower limits are set by a $t\lesssim4$\,hr deceleration time, and the upper limits by a $t\gtrsim1$\,hr deceleration time.

It is important to note that a peak in the light curve can occur earlier or later than the deceleration time owing to other effects.  An earlier peak can be produced by internal-shock processes (flaring), while a later peak can occur due to late-time energy reinjection from the central engine into the external shock or  to the passage of peak synchrotron frequency $\nu_m$ through the optical band.  The former case would void the lower limit, while the latter case would void the upper limit.  The smooth nature of the TESS light curve suggests that the lower limit is probably robust, but the upper limit can certainly be called into question: our modeling (\S \ref{sec:modeling2})  does indeed suggest that $\nu_m$ is likely to be close to the optical band at 1--4\,hr.

\subsubsection{Lower limit from nonthermal spectrum}
\label{sec:selfabslimit}

Compact radio sources exhibit steep radio spectra ($\nu^2$ to $\nu^{5/2}$) on account of synchrotron self-absorption of radio emission from within the dense, shocked gas.
Our first radio observation, at $t=2.86$\,d, may be self-absorbed: it falls below the projected synchrotron spectrum and exhibits a steep downturn toward lower frequencies.  As strong scintillation was occurring at this time and only X-band observations are available, it is not possible to confirm this.  However, by the time of the first multireceiver observation at $t=11$\,d, the radio spectrum is clearly not self-absorbed, indicating it has expanded sufficiently to be optically thin above $\nu>3$\,GHz. 
Using Equation 5 from \cite{BarniolDuran2013} and assuming $\nu_m \approx 50$\,GHz, $\nu_a \leq \nu_m$, and a full filling factor, we estimate a minimum average Lorentz factor of $\Gamma_{\rm av,10d} > 2.7$ (at 10\,d).

This is only an average limit out to late times.  However, the jet velocity is not constant during this phase: the Lorentz factor drops with time as $\Gamma \propto t^{-3/8}$ in a constant-density medium, or as $\Gamma \propto t^{-1/4}$ in a wind medium.
Extrapolating back to the upper limit on the peak time of the afterglow at $\sim 4$\,hr, we infer $\Gamma_{\rm av, 4\,hr} \gtrsim 13$ (uniform) or $\Gamma_{\rm av, 4\,hr} \gtrsim 8$ (wind).

If the first radio epoch was in fact self-absorbed, the equivalent {\it maximum} average Lorentz factor extrapolated to 4\,hr is $\Gamma_{\rm av,4\,hr} \lesssim 11$ in the uniform case, or $\Gamma_{\rm av,4\,hr} \lesssim 8$ in the wind case.  As this is in tension with the more secure estimate from the 11\,d spectrum, this suggests the first epoch was probably not self-absorbed.  (Indeed, a change from self-absorbed to unabsorbed on these timescales would be inconsistent with an afterglow expanding into a constant-density ISM to begin with, as $\nu_a$ is constant within the model.)

\subsubsection{Upper limit from interstellar scintillation}
\label{sec:scintillation}

The radio spectrum (Figure \ref{fig:radioseds}) shows wiggles in frequency space and short-timescale fluctuations (by a factor of $\sim 2$) until at least 30\,days, and probably as late as 130\,days.  This strongly suggests that the source is small enough in angular size until at least 30\,d to be affected by strong interstellar scintillation (ISS) from electrons along the line of sight through our Galaxy.  From Figures 1--2 of \cite{Walker2001} (Erratum to \citealt{Walker1998}), the critical frequency for ISS in this direction is $\nu_0 \approx 12$\,GHz and the Fresnel scale at this frequency is $\theta_{F0} \approx 2.5$\,$\mu$arcsec; the latter corresponds to a physical scale of $6.5 \times 10^{16}$\,cm (25\,light-days) given the angular diameter distance of the source.  Large-amplitude ISS (modulation index $\sim 1$) near $\nu_0$ requires a source size comparable to the Fresnel scale,  
so the implied $\Gamma_{\rm av,30\,d}$ is $\Gamma \leq $\ 2. 

To convert this limit on the average Lorentz factor to a limit on the post-deceleration Lorentz factor, we use the same general reasoning as in \S \ref{sec:selfabslimit} and extrapolate back our late-time limit to the peak time of the afterglow.
However, as our modeling (\S \ref{sec:modeling2}) indicates that a jet break likely took place at 10--20 days, we must consider the post-jet evolution.  Conservatively adopting the earliest possible jet-break time of $t_j = 10$\,d, 
we first extrapolate the 30\,d size constraint to the jet-break time assuming $\theta \propto t^{1/4}$ (the angular size evolution after the jet break; \citealt{Granot+2005}) to obtain $\theta_{10\,d} < 1.9$\,$\mu$as.   We further extrapolate the size evolution from the jet break time to the peak time according to $\theta \propto t^{5/8}$ (ISM) or $\theta \propto t^{3/4}$ (wind).  Finally, we estimate the minimum average Lorentz factor $\Gamma = R / (c\,t_{\rm rest}) = \theta D_{\rm ang}(1+z)/(c\,t_{\rm obs})$.  Conservatively adopting 1\,hr as the time of peak,
we infer an Lorentz factor upper limit of $\Gamma_{\rm av, 1h} < $\ 33 (uniform) or $\Gamma_{\rm av, 1hr} < $\ 17 (wind).

As was the case with the deceleration constraint itself, this limit can be treated as a limit on the true initial Lorentz factor only if the optical peak is due to deceleration. If the observed peak originates from a different mechanism (e.g., $\nu_m$ break), then deceleration occurred earlier and the initial Lorentz factor can be higher.
Additionally, caution is warranted in interpreting constraints based on scintillation arguments, since many of the best-observed GRB afterglows in the literature do not conform well to the predictions of scintillation theory \citep{Alexander+2019,Marongiu+2022}.

\vspace{0.3cm}

Taken together, the three lines of argument above suggest that the bulk of the material along our line of sight was at least moderately relativistic ($\Gamma > 10$), but need not have been highly relativistic ($\Gamma > 50$).

\subsection{Constraints from Afterglow Modeling}
\label{sec:modeling2}

\begin{figure}
\includegraphics[width=0.47\textwidth]{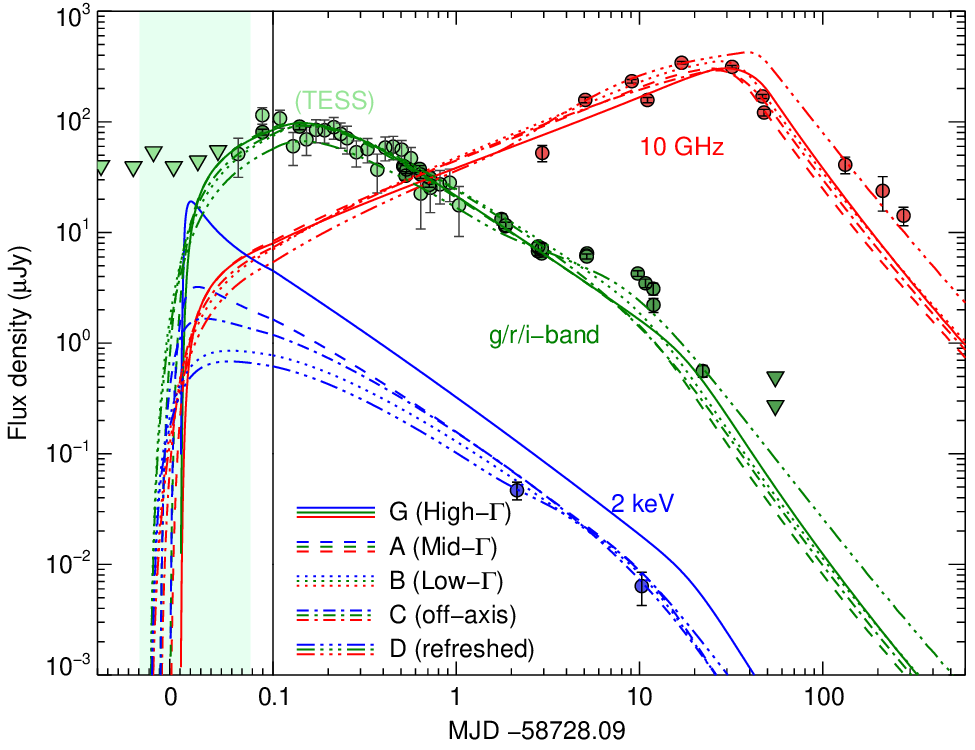}
\caption{Afterglow models fit to the X-ray, optical, and radio light curves of AT\,2019pim.   The five models are described in the text.  All of the models reproduce the basic qualitative behaviour at each wavelength, although none can reproduce the features in detail.  The axis scale is logarithmic in time in the right segment of the plot and linear in the left segment; the observed $g$ and $i$ bands have been shifted to match the $r$ band.}
\label{fig:modelplot}
\end{figure}

Additionally, we modeled the entire afterglow dataset using a numerical code based on the method presented by \cite{Lamb2018}.  Free parameters in the model include the initial Lorentz factor $\Gamma_0$, jet half-opening angle $\theta_j$, viewing angle $\iota$, as well as the (isotropic-equivalent) kinetic energy $E_K$, the circumburst density $n$, and the time of explosion $t_0$ (measured relative to MJD 58728.09).  
This model also permits a variety of jet-structure profiles and allows for the possibility of late-time energy injection \citep[``refreshed'' shocks; see][for details]{Lamb2020}.  We attempted three types of model:  a simple uniform (``top-hat'') jet with no energy injection, a uniform jet with energy injection, and a structured jet without energy injection.  In each case we fix the microphysical parameters $\epsilon_B = 0.001$, $\epsilon_e = 0.1$, $\epsilon_N = 0.15$, and $p = 2.3$ (the fraction of energy given to the magnetic fields, the accelerated electrons, the fraction of accelerated electrons that contribute to synchrotron emission, and the power-law index for the relativistic electron distribution, respectively).  These values were chosen following preliminary exploration of the theoretical parameter space via nested sampling with priors informed by precedent from fitting previous GRBs, as they were able to reproduce the salient features of the data across a variety of models.

Consistent with our analysis using basic physical arguments, most models converge toward Lorentz factors that are lower than typical for GRBs but still relativistic ($\Gamma = 30$--50).  In the case of a structured jet, higher core Lorentz factors are preferred, but the viewing angle is at the edge of the jet core and the material ejected toward the viewer, consistent with a scenario in which the most relativistic material is beamed outside the line of sight.

To contrast various potential interpretations of the afterglow, we focus on five specific cases below:

\begin{itemize}
  \item Model G (``high-$\Gamma$''): A uniform, on-axis jet with a ``high'' Lorentz factor ($\Gamma \approx 100$).  This model is generally expected to produce observable gamma-rays, though an underluminous/soft burst may be possible if the outflow is very smooth \citep{Barraud+2005,Zitouni+2008}.
  \item Model A (``mid-$\Gamma$''): A uniform, on-axis jet with a ``moderate'' Lorentz factor ($\Gamma \approx 55$), close to the threshold where high-energy emission should be suppressed given typical inferred emission region sizes from previous luminous GRBs (see, e.g., \citealt{Lamb+2016,Matsumoto+2019}).
  \item Model B (``low-$\Gamma$''): A uniform, on-axis jet with a ``low'' Lorentz factor ($\Gamma \approx 35$), for which pair production should almost completely suppress high-energy photons.
  \item Model C (``grazing''): A structured relativistic jet with a ``high'' Lorentz factor core, but viewed from just outside this core.
  \item Model D (``refreshed''): A uniform, on-axis, jet with a low to moderate Lorentz factor and late-time energy injection.
\end{itemize}

Model parameters in each case were chosen from regions of the posterior distribution of the model runs that indicate reasonably good fits to the data (at least in comparison to other models); values are given in Table \ref{tab:modelparameters}.  
In the case of model C, $E_K$ is 
the isotropic-equivalent energy along the symmetry axis, and $\theta_j$ refers to the jet structure core angle.  We assume a Gaussian jet profile\footnote{Our choice of a Gaussian for this model is ad-hoc, and some studies have preferred other forms of the dependence of the energy on lateral angle, in particular a power-law: \citep{Beniamini+2022,OConnor+2023,Gill+2023}.  However, the effect of the details of the jet structure on the light curve after peak is limited, and similar conclusions would have been obtained for an alternative structure model.} where the energy and the Lorentz factor vary with lateral angle as 
$E(\theta \leq \theta_e$) = $E_K\,e^{-0.5(\theta/\theta_j)^2}$, and $\Gamma(\theta) = 1 + (\Gamma_0 - 1)\,e^{-0.5(\theta/\theta_j)^2}$
for $\theta < \theta_e$ and $E(\theta > \theta_e) = 0$. In the case of model D, the Lorentz factor of the decelerating blast wave when energy injection begins is given by $\Gamma_c$ and the fractional energy increase originating from the refreshed shock is parameterised as $f_e$. 

The model light curves are plotted against the data in Figure \ref{fig:modelplot}.  The $g$ and $r$ optical bands have been shifted to match the $i$ band, as have the TESS data.

\setlength{\tabcolsep}{4pt}
\begin{table}
	\centering
	\caption{Model parameters}
	\label{tab:modelparameters}
	\begin{tabular}{lllllll}
		\hline
		&  & G & A & B & C & D \\
		\multicolumn{2}{l}{Parameter} & high-$\Gamma$ & mid-$\Gamma$ & low-$\Gamma$ & off-axis & refreshed \\
		\hline
         $\Gamma_0$ [$\Gamma_c$]&           & 100  &  55    & 35     & 130      & 45 [7]    \\
         $E_K$ [$f_e$]    &(10$^{53}$\,erg) & 3.5  &  2     & 2      & 6        & 1.6 [6.4] \\
         $n$       & (cm$^{-3}$)            & 0.5  &  2     & 3      & 2.5      & 2.5       \\
         $\theta_j$ [$\theta_e$]& (rad)     & 0.22 &  0.22  & 0.22   & 0.09 [0.4]& 0.15   \\
         $\theta_\iota$   & (rad)           & 0    &  0     & 0      & 0.13      & 0       \\ 
         $t_0$     & (day)                  & 0.01 &  0     & -0.02  & -0.01     & -0.02    \\
         \hline
	\end{tabular}
\newline 
\end{table}

It can be seen that all of these models reproduce the basic observations (the approximate peak times, decay slopes, and relative fluxes in each band), although none of them fully reproduce the optical flattening or the much steeper evolution in the optical compared to the radio at late times.  
Model D (``refreshed'') comes closest to reproducing the late-time evolution (this model was introduced for this reason), though it does not fully explain the optical bump feature and it overpredicts the radio data around peak brightness.  Model A (mid-$\Gamma$) underpredicts both the late-time optical and radio data but better explains the rise timescale.  Model G (high-$\Gamma$) is similar but also greatly overpredicts the X-rays.

We cannot formally rule out any of the scenarios on the basis of the afterglow alone, both owing to the simplified nature of the models and because we have not yet performed an exhaustive search of the parameter space for each case.  However, the modeling establishes that a low-$\Gamma$ outflow is indeed consistent with most of the key features in the data (rise time, decay rate, and multiwavelength spectrum). On the other hand, while a high-$\Gamma$ on-axis outflow is not a good match to the data, a classical GRB is fully consistent with the observations if the jet was observed slightly off-axis.

\section{Conclusions}
\label{sec:conclusions}

While AT\,2019pim is unambiguously the afterglow of a relativistic explosion, its rise time to peak is substantially slower than is typical of afterglows of known gamma-ray bursts, and nondetections by GBM and Konus rule out prompt gamma-ray emission at a limit comparable to the fluence expected for GRB afterglows of comparable luminosity.  These properties are consistent with, although do not strictly require, a model in which the early afterglow radiation that is observed is produced by ejecta moving toward us at a moderate initial Lorentz factor ($\Gamma \approx$\ 10--50).  This is substantially less than what has been reported for any classical (i.e., non-low-luminosity) GRB to date (\citealt{Ghirlanda+2018,Chen+2018,Hascoet+2014}, although c.f. \citealt{DereliBegue+2022}.)

Our data are not able to distinguish between models under which the low-$\Gamma$ material originates from an on-axis jet with an intrinsically low initial Lorentz factor (a ``dirty fireball''), versus low-$\Gamma$ material from the high-latitude component of a structured jet seen partially off-axis (such that only the material along our line of sight is low-$\Gamma$, and a classical GRB was produced in some other direction).  Additionally, while our modeling does not prefer a high-$\Gamma$ on-axis scenario, we cannot strictly rule out a scenario in which AT2019pim is the afterglow of a GRB with low gamma-ray efficiency, particularly if it occurred during one of the {\it Fermi} occultations.

Additional intensive studies of future ``orphan'' afterglow events will be needed to securely identify whether dirty fireballs truly exist in nature (or to rule them out if they do not), and to study the structure of the jet in classical GRBs.
Fortunately, AT\,2019pim was only the first example of a well-observed optically selected afterglow of this nature.  Since the discovery of AT\,2019pim, ZTF has already increased the size of the orphan afterglow sample by an order of magnitude, including the discovery of AT\,2021lfa, which also shows compelling evidence of an extended ($\sim$3 hr) rise time characteristic of a dirty fireball \citep{Lipunov2022}, and the recent detection of an afterglow with even more constraining limits on accompanying gamma-ray emission (Li et al. 2024, in prep.)
Particularly notable is the detection of AT\,2022cmc, a relativistic transient with an inferred Lorentz factor of only $\Gamma \approx 10$ (\citealt{Andreoni+2022,Pasham+2023,Rhodes+2023}, c.f. \citealt{Yao+2023}):
while its origin appears to be due to a tidal disruption of a star rather than a collapse, it clearly demonstrates that optical surveys are quite sensitive to energetic relativistic transients across the entire range of potential Lorentz factors.  While the assemblage of optically selected afterglows remains too small at the present time to draw firm conclusions on the nature of the population, the techniques to find events of this nature are now well-established and should lead to more discoveries in the coming years.

Continued improvements to afterglow search methods in large surveys and the commissioning or expansion of additional wide-field facilities capable of high-cadence monitoring over large areas (such as GOTO and ATLAS) should increase the discovery rate in the coming years, and even more powerful surveys such as the upcoming Large Array Survey Telescope \citep{Ofek+2023} and proposed Argus Array \citep{Law+2022} will further extend these capabilities.  Additionally, upcoming powerful radio facilities such as the Next-Generation VLA (ngVLA; \citealt{Murphy+2018}) and Square Kilometre Array (SKA) will enable late-time calorimetry and possibly direct imaging of the jet, permitting distinguishing off-axis from on-axis cases.  Soft X-ray surveys (including the recently launched Einstein Probe, and proposed future facilities such as THESEUS or HiZ-GUNDAM; \citealt{Amati+2021,Ghirlanda+2021,Yonetoku+2014}) also represent a promising means of low-$\Gamma$ afterglow discovery.
Comprehensive observational studies of individual events gathered by each of these surveys, together with comparative studies of afterglow populations selected at different wavelengths and different timescales, will allow us to finally produce a complete picture of energetic, relativistic ejection from collapsing stars.

\section{Data Availability}

All flux density measurements of the afterglow underlying this article are available in the article and in its online supplementary material.  Additionally, the original raw data from LT, Keck, TESS, and VLA are available in public archives hosted by their respective facilities.

\section{Acknowledgments}

We thank the anonymous referees for valuable feedback on the draft of this paper.  We also thank S. Kobayashi for advice and input.

This work was supported by the GROWTH project funded by the U.S. National Science Foundation (NSF) under grant 1545949. GROWTH is a collaborative project between California Institute of Technology, Pomona College, San Diego State University, Los Alamos National Laboratory, University of Maryland College Park, University of Wisconsin Milwaukee (USA), Tokyo Institute of Technology (Japan), National Central University (Taiwan), Indian Institute of Astrophysics (India), Weizmann Institute of Science (Israel), The Oskar Klein Centre at Stockholm University (Sweden), Humboldt University (Germany), and Liverpool John Moores University (UK).  This paper used the GROWTH marshal to filter alerts and coordinate follow-up observations.

The discovery, early follow-up observations, and initial analysis of this transient took place during the Aspen Center for Physics (ACP) summer 2019 workshops, ``Astrophysics in the LIGO/Virgo Era'' and ``Non-standard Cosmological Probes''.  DAP thanks the ACP for hospitality and support during this period, and acknowledges fruitful discussions with P. Chandra, J. Granot, K. Alexander, and many others on this event.   The ACP is supported by NSF grant PHY-1607611.  This work was partially supported by a grant from the Simons Foundation.

AYQH was supported by an NSF Graduate Research Fellowship under grant DGE-1144469.
GPL is supported by a Royal Society Dorothy Hodgkin Fellowship (grants DHF-R1-221175 and DHF-ERE-221005).
S.~Anand acknowledges support from NSF GROWTH PIRE grant 1545949.
AC acknowledges support from NSF grant AST-2307358.
ECK acknowledges support from the G.R.E.A.T research environment, funded by Vetenskapsrådet, the Swedish Research Council, under project 2016-06012.
RH acknowledges funding from the European Union’s Horizon 2020 research and innovation programme under Marie Skłodowska-Curie grant agreement 945298-ParisRegionFP.
AVF's team at UC Berkeley is grateful for financial support from the Christopher R. Redlich Fund, William Draper, Timothy and Melissa Draper, Briggs and Kathleen Wood, Sanford Robertson (TGB is a Draper-Wood-Robertson Specialist in Astronomy), and many other donors.
PGJ has received funding from the European Research Council (ERC) under the
European Union’s Horizon 2020 research and innovation programme (Grant agreement No.~101095973).

Based on observations obtained with the Samuel Oschin 48-inch telescope and the 60-inch telescope at the Palomar Observatory as part of the Zwicky Transient Facility project. ZTF is supported by the NSF under grant AST-1440341 and a collaboration including Caltech, IPAC, the Weizmann Institute of Science, the Oskar Klein Center at Stockholm University, the University of Maryland, the University of Washington, Deutsches Elektronen-Synchrotron and Humboldt University, Los Alamos National Laboratories, the TANGO Consortium of Taiwan, the University of Wisconsin at Milwaukee, and Lawrence Berkeley National Laboratories. Operations are conducted by COO, IPAC, and UW.

The Karl G. Jansky Very Large Array is operated by NRAO, for the NSF under cooperative agreement by Associated Universities, Inc.
This paper includes data collected by the TESS mission. Funding for the TESS mission is provided by the NASA Explorer Program.
This research is based in part on observations made with the {\it Neil Gehrels Swift Observatory}. 

The Liverpool Telescope is operated on the island of La Palma by Liverpool John Moores University in the Spanish Observatorio del Roque de los Muchachos of the Instituto de Astrofisica de Canarias with financial support from the UK Science and Technology Facilities Council.

Some of the data presented herein were obtained at the W. M. Keck Observatory, which is operated as a scientific partnership among the California Institute of Technology, the University of California, and the National Aeronautics and Space Administration (NASA). The Observatory was made possible by the generous financial support of the W. M. Keck Foundation. The authors wish to recognise and acknowledge the very significant cultural role and reverence that the summit of Maunakea has always had within the indigenous Hawaiian community.  We are most fortunate to have the opportunity to conduct observations from this mountain.  We thank Weikang Zheng for assistance with obtaining the late-time Keck imaging.

The GROWTH-India telescope (GIT; Kumar et al 2022) is a 70 cm telescope with a 0.7 deg$^2$ field of view set up by the Indian Institute of Astrophysics (IIA) and the Indian Institute of Technology Bombay (IITB) with funding from the Indo-US Science and Technology Forum and the Science and Engineering Research Board, Department of Science and Technology, Government of India. It is located at the Indian Astronomical Observatory (IAO, Hanle). We acknowledge funding by the IITB alumni batch of 1994, which partially supports the operation of the telescope. Telescope technical details are available at https://sites.google.com/view/growthindia/. 

The Pan-STARRS1 Surveys (PS1) and the PS1 public science archive have been made possible through contributions by the Institute for Astronomy, the University of Hawaii, the Pan-STARRS Project Office, the Max-Planck Society and its participating institutes, the Max Planck Institute for Astronomy, Heidelberg and the Max Planck Institute for Extraterrestrial Physics, Garching, The Johns Hopkins University, Durham University, the University of Edinburgh, the Queen's University Belfast, the Harvard-Smithsonian Center for Astrophysics, the Las Cumbres Observatory Global Telescope Network Incorporated, the National Central University of Taiwan, the Space Telescope Science Institute, the National Aeronautics and Space Administration under Grant No. NNX08AR22G issued through the Planetary Science Division of the NASA Science Mission Directorate, NSF grant AST-1238877, the University of Maryland, Eotvos Lorand University (ELTE), the Los Alamos National Laboratory, and the Gordon and Betty Moore Foundation.

\bibliographystyle{mnras}
\bibliography{ref}

\end{document}